\documentclass[journal=jpcbfk,manuscript=article]{achemso}

\usepackage{textcomp}
\usepackage{longtable}
\usepackage{multirow}
\usepackage{amsmath,amssymb,wasysym}
\usepackage{amsfonts,mathtools}
\usepackage{subfigure}
\usepackage[usenames,dvipsnames]{color}
\usepackage{rotating}
\usepackage{threeparttable}
\usepackage{xr}

\usepackage{dcolumn}            %
\usepackage{hyperref}           %
\usepackage{cleveref}
\usepackage{IEEEtrantools}      %
\usepackage{wasysym,textcomp}   %
\usepackage[version=4,arrows=pgf]{mhchem}
\usepackage{chemfig}
\usepackage{siunitx}            %
\usepackage{pgf,tikz}     %
\usepackage{booktabs,makecell}  %

\usepackage{savesym}
\savesymbol{fax}
\usepackage{marvosym}
\restoresymbol{Marvo}{fax}

\newcolumntype{d}[1]{D{.}{\cdot}{#1}} %

\DeclareSIUnit[number-unit-product = {\,}]
\cal{cal}
\newcommand\mathplus{+}
\usepackage{xr}
\author{Luis Vasquez}
\affiliation{Institute of Applied Radiation Chemistry, Lodz University of Technology,
{\.Z}eromskiego 116, {\L}{\'{o}}d{\'z} 90-914, Poland}
\email{luis.vasquez@edu.p.lodz.pl}

\title {Partial Hessian vibrational analysis: vapour pressure isotope effects and their relation with non-covalent interactions}

\begin{document}

\clearpage
\date{\today}

\begin{abstract}

Stable isotope analysis is regularly used in order to acquire detailed information of physical and chemical processes that a system is undergoing. Experimental data is thus complemented with computational studies, in order to develop theoretical models that can satisfactorily explain and predict a determined phenomenon in similar systems. However, proper estimation of isotope effects remains a challenging task. High levels of theory are demanded in order to correctly describe non-covalent interactions, while large molecular systems are required to mimic solvent effects. Given specifications typically restrict the available theoretical approach to a few electronic structure methods.

In this study equilibrium isotope effects (EIEs) on evaporation to several organic solvents (bromobenzene, dibromomethane, ethanol, methanol, and trichloromethane) in the pure phase are estimated employing Kohn-Sham Density functional theory (KS-DFT) along with full Hessian vibrational analysis (FHVA) and partial Hessian vibrational analysis (PHVA). Both FHVA and PHVA qualitatively predict compounds EIEs. Weakly interacting systems (bromobenzene, dibromomethane, and trichloromethane) display identical EIE values for both, FHVA and PHVA. Whereas for strongly interacting systems (ethanol and methanol), PHVA estimates slightly smaller EIE compared to FHVA. By employing the symmetry-adapted perturbation theory (SAPT) along with independent gradient model (IGM), it was possible to establish how minimal changes in compounds interaction energy affected carbon and bromine EIEs estimation in dibromomethane. Moreover, SAPT and IGM revealed how in ethanol, apparently insignificant intermolecular interactions possessed an enormous impact in carbon position specific isotope effects, which highlight the importance of solvent effects for strongly interacting compounds. The presented results have important implications for correctly comprehending and characterising the role of non-covalent interactions in the prediction of isotope effects, as well as providing a robust test to PHVA for the prediction of EIEs.

\end{abstract}

\section{\label{sec:intro}Introduction}

Equilibrium isotope effects (EIEs) are one of the most successful tools in biochemistry, geochemistry, and chemistry for meticulously studying exchange reactions during processes such as volatilisation, sorption, and evaporation \cite{Julien2017,Julien2015a,Horst2016,Heckel2019}. Given processes are vital to comprehend pollution, biodegradation pathways and remediation of contaminated aquiferous systems like oceans, lakes, and underground reservoirs. Halogenated compounds such as trichloromethane and bromobenzene represent a global concern, given that are not merely perilous pollutants with suspected carcinogen properties, but equally are known for their slow dissolution (per years) in anoxic environments \cite{Heckel2019}.

Technological advances during the last two decades permit now the accurate measuring of position specific isotope effects, hence providing an valuable insight into the nature of bulk isotope effects. However, precise interpretation of experimental results remains a complex task, predominantly owing to physicochemical phenomena that might be easily overlooked such as diffusion or adsorption, but posses a substantial influence over the outcome \cite{Julien2015,Kopinke2017}. For instance, it is conceivable that the diffusion effect vanishes from compounds boiling point on, leaving as result mere EIE or so-called vapour pressure isotope effects (VPIE) \cite{Julien2015a,Jancso1974}. However when direct measurements are performed at standard conditions, to some extend the diffusion effects are present, under those circumstances the term volatilisation isotope effect (VIEs) is employed to emphasise the temperature factor.

Usually, owing to the quantum nature of EIEs \cite{Bigeleisen1947}, experimental data is adequately depicted \emph{via} electronic structure calculations. Thus, finding stationary points on a potential-energy surface (PES), the theory of molecular vibrations of molecular systems \cite{Wolfsberg2009} is employed by the Bigeleisen-Mayer approach for the practical calculation of isotope effects \cite{Bigeleisen1947}. Predominantly, electron structure calculations based on Kohn-Sham density functional theory (KS DFT) \cite{Kohn1965} are used. Given choice is motivated mainly by the relative robust performance of density functionals and their affordable computational cost for the full quantum-mechanical treatment of large systems (molecular systems with more than 40 atoms). In practice, the B3LYP functional \cite{Vosko1980,Lee1988,Becke1993,Stephens1994} is likely the most used functional for isotope effects calculation \cite{Wolfsberg2009}, and although in recent KS DFT benchmarks it does not perform well \cite{Goerigk2017}, its notable ability to predict correctly the isotopic frequency shifts converts it in one of the DFT functionals for excellence. Another exceptional functional is MPW1PW91 \cite{Adamo1999,Ernzerhof1999,Adamo1998}, which has permitted the accurate predictions of hydrogen and carbon VPIEs for fluoroform \cite{Oi2017,Yanase2007}.

Important, and a point often overlooked, is the quality of stationary point atomic positions. It must be remembered that the Hessian is the second derivative of the molecular energy with respect to nuclear coordinates (Cartesian coordinates), therefore undergoing an accurate reproduction of them intentionally serves as a pivotal step in order to calculate reliable isotope effects for the precise reason. Moreover, correctly reproducing compounds interatomic interactions conjointly guarantee solvent effects emulation. Goerigk \emph{et al.} have demonstrated in a recent study that weighted total mean absolute deviation for non-covalent interaction calculated by B3LYP and MPW1PW91 functionals are notably large, especially if dispersion corrections are not included \cite{Goerigk2017}. Therefore, meticulous care during the functional selection is required, given that generally accurate atomic positions implies expensive numerical calculations within the "Jacob’s ladder" rungs.

After carefully choosing an adequate DFT functional for the purpose of calculating EIEs, the Hessian matrix must be computed, however, because the Hessian matrix is routinely calculated numerically (finite difference of analytic gradients) \cite{Pulay1969,McIver1972} within KS DFT, even for relatively compact systems its calculation is computationally expensive \cite{Rahalkar2008}. In order to surmount given drawback, either software package parallelisation along with more efficient algorithms or rigorous Hessian approximations are been continuously developed \cite{Rahalkar2008,Sharada2014,Yamamoto2017,Yang2018,Pascale2004}. The later is appealing, particularly the partial Hessian approach \cite{Head1997,Head1999}, which principle lies on the diagonalisation of a sub-block of the mass-weighted energy second derivative matrix computed for a subgroup of atoms. This pragmatic approach has been successfully employed for characterisation of the infrared spectrum of large transition metals \cite{SaeHeng2018}, vibrational analysis of relevant cluster compounds \cite{Besley2008}, fingerprint vibrations of adsorbates on surfaces \cite{Besley2008,Head1997,Head1999}, and molecular vibrational energy and entropy \cite{Li2002}. Although partial Hessian \cite{Head1997} or adapted versions have been efficiently implemented in the past for estimation of EIEs \cite{Yuan2014,Rustad2008}, those studies did not explore at least one of the following issues i.) the quality of intermolecular interactions compared to experimental data or reliable levels of theory, ii.) equivalent configurations of PES by nuclear coordinates, iii.) effect of partial Hessian in the reduced partition function ratio (RPFR) terms \cite{Bigeleisen1947}, or iv.) practical limitation of partial Hessian approach.

We report on numerical calculation results of EIEs for two organic solvents which mainly interact \emph{via} hydrogen bonds: ethanol, and methanol, and other four highly polluting halogenated compounds bromobenzene, trichlorofluoromethane, dibromomethane, and trichloromethane. EIEs results from full Hessian and partial Hessian vibrational analysis are properly compared, as well as their direct effects on the components of the reduced partition function ratio. Moreover, unique insight into intermolecular interactions of certain studied compounds and their relationship with EIEs are carefully analysed and described.

\section{\label{sec:material} Methods}

Molecular dynamic simulations were performed for bromobenzene (BB - \ce{C6H5Br}), trichlorofluoromethane (CFC - \ce{CFCl3}), dibromomethane (DBM - \ce{CH2Br2}), ethanol (ETH - \ce{C2H5OH}), methanol (MeOH - \ce{CH3OH}), and trichloromethane (TCM - \ce{CHCl3}) (\cref{tab:molecules}) using the open-source software LAMMPS (large-scale atomic/molecular massively parallel simulator) \cite{Plimpton1995}. Optimized potentials for liquid simulations (OPLS) force field parameters \cite{Jorgensen1996,Jorgensen1988} from the LigParGen web server \cite{Dodda2017} have been used for model simulations. Periodic boundary conditions for a cubic box comprising 90 molecules have been properly minimised and brought to equilibrium at standard conditions.

\begin{table*}[htp!]
	\caption{Systems used in this study.}
	\label{tab:molecules}
		\begin{tabular*}{\textwidth}{@{\extracolsep{\fill}}lc}
\toprule
			Name                            & Structure                                        \\
			\midrule
			Bromobenzene (BB)               & \chemfig{Br-[:-150]C_1*6(=C_2-C_3=C_4-C_5=C_6-)} \\
			                                &                                                  \\
			Dibromomethane (DBM)            & \chemfig{Br-[:30]C-[:-30]Br}                      \\
			                                &                                                  \\
			Chloroform (TCM)                & \chemfig{Cl-[:30]C(<[:-60]Cl)<:[:-15]Cl}  \\
			                                &                                                  \\
			Trichlorofluoromethane (CFC) & \chemfig{Cl-[:30]C([2]-F)(<[:-60]Cl)<:[:-15]Cl}     \\
			                                &                                                  \\
			Ethanol (EtOH)                  & \chemfig{C_1-[:30]C_2-[:-30]OH}                  \\
			                                &                                                  \\
			Methanol (MeOH)                 & \chemfig{C-[:00]OH}                		   \\
\bottomrule		
\end{tabular*}
\end{table*}

Consequently, for each compound, five clusters containing four molecules were extracted from trajectories of equilibrated MD simulations. Nuclear coordinates of clusters were further optimised using a long-range-corrected (LRC) hybrid functional with non local correlation, \(\omega\)B97X-V \cite{Mardirossian2014}. Basis set double, and triple Dunning's augmented correlation consistent \cite{Parrish2013,Parker2014} were used. All geometrical optimisations were carefully performed using the SG-2 \cite{Dasgupta2017} integration grid. Stationary points on the potential surface were confirmed from full Hessian vibrational analysis (FHVA) without imaginary frequencies. During the vibrational analysis the SG-3 grid \cite{Dasgupta2017} was employed. All quantum mechanical calculations were carried out in Q-Chem software packages (version 4.4) \cite{Shao2014}, and were typically restricted to an allowable maximum of 4032 computational hours.

Partial Hessian vibrational analyses \cite{Besley2008,Head1997} (PHVA) were performed to clusters for which FHVA was carried out successfull. In PHVA only one molecule was "active", \emph{i.e.} the second partial derivatives of the potential with respect to displacement of the atoms is calculated, whereas the other three remained "frozen". Given approach was used with each one of the molecules within a cluster.

With the purpose to robustly assess the accuracy of PHVA with respect to FHVA, equilibrium isotope effects (EIE) were properly calculated implementing the Bigeleisen-Mayer approximation from the ratio of the free energies of the isotopologues \cite{Bigeleisen1947}, thus, using vibrational analysis for Bigeleisen-Mayer approximation gives
\begin{IEEEeqnarray}{rCl}
     \label{eq:iso_effect_vibrati}
     \frac{s_h}{s_l}f=\prod_i \frac{\nu_{hi}}{\nu_{li}} \frac{exp\left(-hc\nu_{li}/2k_BT\right)/1-exp\left(-hc\nu_{li}/k_BT\right)}{exp\left(-hc\nu_{hi}/2k_BT\right)/1-exp\left(-hc\nu_{hi}/k_BT\right)} = EIE
\end{IEEEeqnarray}
where \(h\) is the Planck's constant, \(c\) represents the speed of light in vacuum, \(k_B\) introduces the Boltzmann constant, \(T\) refers to temperature, \(\nu\) indicates the harmonic frequency, and \(\left({s_h}/{s_l}\right)f\) is the reduced partition function ratio.

It is worth to note that EIE can be expressed in terms of enrichment factor \(\varepsilon\) as
\begin{equation}
	\label{eq:enr-factor}
	\varepsilon = \frac{1}{n} \left(\frac{1}{EIE} -1 \right) \cdot 1000  ,
\end{equation}
where \(n\) represents the number of atoms of the same element in a molecule (\cref{sec:material}) that have been isotopically substituted. EIE and \(\varepsilon\) have been calculated in the developing version of the open source program isoC \cite{Vasquez2019}. Enrichment factor \(\varepsilon\) is used along this work for consistency with experimental data.

A decomposition energy analysis is performed in order to obtain an insight into compounds interactions strength and correctly identify patters between FHVA and isotope effects. Clusters were divided in single molecules (four per cluster) and all arranges (six) of possible dimers were promptly formed, consequently, interaction energy \(E_{\textrm{int}}\) for dimers were calculated using the symmetry-adapted perturbation theory (SAPT) \cite{Jeziorski1994}. Within the so-called supermolecular approach that SAPT follows, interaction energy \(E_{\textrm{int}}\) of a system typically consisting of two separable subsystems \(A\) and \(B\) is defined as
\begin{equation}
    \label{eq:raw_interact_energy}
    E_{\textrm{int}} = E_{\textrm{AB}}- E_{\textrm{A}}- E_{\textrm{B}}.
\end{equation}
In parallel, the SAPT interaction energy can be further expanded and decomposed into four physically meaningful components \emph{i.e.} electrostatic, exchange, induction, and dispersion yielding 
\begin{equation}
    \label{eq:interact_energy_four_term}
    E_{\textrm{int}} = E_{\textrm{elst}}+ E_{\textrm{exch}}+ E_{\textrm{ind}} + E_{\textrm{disp}},
\end{equation}
where \(E_{\textrm{elst}}\) is the electrostatic component, \(E_{\textrm{exch}}\) represents the exchange term, \(E_{\textrm{ind}}\) gives the inductions component, and \(E_{\textrm{disp}}\) indicates the dispersion term of interaction energy. Owing to the relative compact size of the studied molecules, a large basis set (Dunning's correlation-consistent augmented triple) along with accurate higher-order SAPT (SAPT+2) with coupled-cluster doubles (CCD) treatment of dispersion and supermolecular MP2 interaction energy (\(\delta\)MP2) \cite{Parrish2013,Parker2014} is utilised for \(E_{\textrm{int}}\) accurate calculation. Interaction energies were calculated in the standard release of PSI4 software package (version 1.3.2) \cite{Parrish2017}.

Complementarity to SAPT analysis, non-covalent interactions were visualised and meticulously studied by means of independent gradient model (IGM) \cite{Lefebvre2018}. Intermolecular interactions described by the density gradient as implemented in the stable release of Multiwfn package software (version 3.6) was used \cite{Lu2011}.

\section{\label{sec:results} Results and discussion}

\Cref{tab:rmsd} contains average root-mean-square deviation (RMDS) of aligned atomic positions from successfully optimised geometries that also were confirmed to be stationary points without negative frequencies.
\begin{table*}
    \caption{Root-mean-square deviation of atomic positions for geometrically optimised clusters (CFC, DBM, EtOH, MeOH, and TCM) at \(\omega\)B97X-V/aug-cc-pvdz and \(\omega\)B97X-V/aug-cc-pvdz levels of theory.}
	\label{tab:rmsd}
	
		\begin{tabular*}{\textwidth}{@{\extracolsep{\fill}}lc		
            d{1.3} d{1.3}}
\toprule
                     &                & \multicolumn{2}{c}{RMSD [\si{\angstrom}]}      \\
\cmidrule(lr){3-4}
            Compound & Atoms compared & \multicolumn{1}{c}{aug-cc-pvdz}  &  \multicolumn{1}{c}{aug-cc-pvtz}      \\
			\midrule
			CFC      & all  &     2.41        & 2.24 \\
			CFC      & C    &     1.82        & 1.85 \\
			DBM      & all  &     1.94        & 1.87 \\
            DBM      & C    &     1.51        & 1.46 \\
            DBM      & Br   &     2.13        & 2.14 \\
			EtOH     & all  &     1.95        & 2.00 \\
			EtOH     & C    &     1.74        & 1.78 \\
			EtOH     & O    &     1.62        & 1.39 \\
			MeOH     & all  &     1.90        & 1.87 \\
			MeOH     & C    &     1.74        & 1.73 \\
			MeOH     & O    &     1.33        & 1.27 \\
			TCM      & all  &     2.00        & 1.90 \\
			TCM      & C    &     1.48        & 1.34 \\
\bottomrule
		\end{tabular*}
	
\end{table*}
Although all five BB clusters were geometrically optimised, stationary points could not be validated from FHVA calculations as those exceeded the computational time limit. It is important to realise that generally speaking, clusters are heterogeneous and possess relatively large RMSD, indicating that VIE and VPIE predictions are obtained to several configurations of the PES\@. The increase of the basis from aug-cc-pvdz to aug-cc-pvtz, in general only slightly decrease the RMSD.

Two pair of clusters, ones in case of EtOH (EtOH-1 and EtOH-2) and MeOH (MeOH-3 and MeOH-4), resulted in identical atomic positions, which can be interpreted as likely abundant configuration within the PES\@. Another key point from RMSD is its minimal value for oxygen atoms in EtOH and MeOH, probably owing to their relevance in the formation of hydrogen bonds (HB). It is well recognised that given compounds mainly interact \emph{via} hydrogen bonding, a weak non-covalent interaction, thus it is reasonably expected that oxygen atomic positions would not variate or at least not as halogen atoms, which non-covalent interactions are significantly weaker.

Average VIEs for studied compounds are given in \cref{tab:vapie_dz}.
\begin{table*}
	\caption{VIEs ($\varepsilon\enspace\permil$) predicted at standard conditions for trichlorofluoromethane, dibromomethane, ethanol, methanol, and trichloromethane, at \(\omega\)B97X-V/aug-cc-pvdz level of theory.}
	\label{tab:vapie_dz}
		\begin{tabular*}{\textwidth}{@{\extracolsep{\fill}}lc	
        d{1.3} r}%
\toprule
            Compound & isotope        &  {\varepsilon}   &  \multicolumn{1}{c}{Exp.}      \\
			\midrule
			CFC      & \ce{^13 C}     &     0.64         &  {1.7\footnotemark[1]}          \\
			DBM      & \ce{^81 Br}    &    -1.74         & {-0.8$\pm$0.1\footnotemark[2]}  \\
            DBM      & \ce{^13 C}     &    -1.99         &  {0.6$\pm$0.2\footnotemark[2]}  \\
			EtOH     & \ce{^13 C_{1}} &    -1.26         &  -6.1\footnotemark[3]           \\
			EtOH     & \ce{^13 C_{2}} &     0.56         &  -2.6\footnotemark[3]           \\
			MeOH     & \ce{^13 C}     &     1.04         &  -6.5\footnotemark[3]           \\
			TCM      & \ce{^13 C}     &     0.16         &   1.2\footnotemark[1]           \\
\bottomrule
		\end{tabular*}
	\raggedright
	\footnotemark[1]{Ref. \citenum{Horst2016}.}   \\
	\footnotemark[2]{Ref. \citenum{Vasquez2018}.} \\
	\footnotemark[3]{Ref. \citenum{Julien2015} by passive evaporation.} \\
\end{table*}
Isotope effect results for CFC produce a qualitatively match with respect to experimental data \cite{Horst2016}, however it is underestimated. Small mean standard deviation (STD) from position-specific isotope effects (see \cref{tab:vapie_dz_sup} of supplementary information) in CFC along with RMSD, indicate that although cluster atomic positions differ appreciably from each other, this variety in the PES does not affect the final prediction.

\ce{^81 Br} isotope effect for DBM is overestimated, while \ce{^13 C} is rightly predicted as normal, but its magnitude is three timer larger than the one from experimental data. The last from the group of halogenated compounds is TCM, which \ce{^13 C} isotope effect is only predicted qualitatively, \emph{i.e.} an inverse isotope effects is predicted, however its value is a factor of magnitude underestimated.

Average position-specific isotope effect for \ce{^13 C_{1}} in EtOH is underestimated, whereas \ce{^13 C_{2}} is not predicted qualitatively either quantitatively. Nevertheless, it is worth mentioning that position-specific isotope effects values for \ce{^13 C_{2}} were in their vast majority close to zero, with only a couple being larger than one. Thus, if \ce{^13 C_{2}} is recalculated with the position-specific values closer to zero, the predicted isotope effect respectively becomes 0.27\(\si{\permil}\). The apparent reason behind such particularly differences in the positions-specific magnitudes is diligently studied and explained in the next section (\cref{sec:sapt}).

According to our compounds scheme, the second carbon position in EtOH and the carbon position in MeOH are similar, as both atoms are bonded to \ce{OH}, therefore it is interesting to characterise how different or similar their position-specific isotope effects are. Predicted \ce{^13 C} in MeOH is a moderate inverse effect (1.04\(\si{\permil}\)), contrary to the considerable normal effect observed in the experimental data. The STD for \ce{^13 C} in MeOH reveals that computed VIE fluctuates along a broad range of values (\(\pm\)3.9), therefore, first it is necessary to minimise the STD in order to infer further conclusions.

The expansion of the basis set to aug-cc-pvtz for the calculation of VIEs provide a drastic change in the predicted values of DBM and MeOH (\cref{tab:vapie_tz}), the rest of the compounds did not undergo relevant changes. In general, \ce{^81 Br} and \ce{^13 C} isotope effects for DBM were diminished at least by a factor of magnitude, thus \ce{^81 Br} passed from -1.17 to -0.02\(\si{\permil}\) and \ce{^13 C} from -1.99 to -0.17\(\si{\permil}\). For \ce{^13 C} in MeOH, it is worth noting that although the calculated isotope effect agrees qualitatively with experimental results, \emph{i.e.} predicted value reflects a normal isotope effect, its magnitude is severely underestimated. Additionally, it is observed that in general the STDs (see \cref{tab:vapie_tz_sup} of supplementary information) have became smaller, especially in the case of MeOH.

\begin{table*}
	\caption{VIEs ($\varepsilon\enspace\permil$) predicted at standard conditions for trichlorofluoromethane, dibromomethane, ethanol, methanol, and trichloromethane, at \(\omega\)B97X-V/aug-cc-pvtz level of theory.}
	\label{tab:vapie_tz}
		\begin{tabular*}{\textwidth}{@{\extracolsep{\fill}}lc		
        d{1.3} r}%
\toprule
            Compound & isotope        &  {\varepsilon}   &  \multicolumn{1}{c}{Exp.}     \\
			\midrule
			CFC      & \ce{^12 C}     &    0.64          & {1.7\footnotemark[1]} \\
			DBM      & \ce{^81 Br}    &   -0.02          &{-0.8$\pm$0.1\footnotemark[2]}  \\
            DBM      & \ce{^13 C}     &   -0.17          & {0.6$\pm$0.2\footnotemark[2]}  \\
			EtOH     & \ce{^13 C_{1}} &   -1.18          & -6.1\footnotemark[3]           \\
			EtOH     & \ce{^13 C_{2}} &    0.47          & -2.6\footnotemark[3]           \\
			MeOH     & \ce{^13 C}     &   -0.13          & -6.5\footnotemark[3]           \\
			TCM      & \ce{^13 C}     &    0.09          &  1.2\footnotemark[1]           \\
\bottomrule
		\end{tabular*}
	\raggedright
	\footnotemark[1]{Ref. \citenum{Horst2016}.}   \\
	\footnotemark[2]{Ref. \citenum{Vasquez2018}.} \\
	\footnotemark[3]{Ref. \citenum{Julien2015} by passive evaporation.} \\
\end{table*}

In conjunction with VIEs, VPIE were calculated for compounds with present experimental data. \ce{^13 C} VPIE for EtOH and MeOH at \(\omega\)B97X-V/aug-cc-pvdz and \(\omega\)B97X-V/aug-cc-pvtz levels of theory are given in the \cref{tab:vpie_dz,tab:vpie_tz}, respectively.
\begin{table*}
	\caption{VPIEs ($\varepsilon\enspace\permil$) for ethanol, and methanol, predicted at \(\omega\)B97X-V/aug-cc-pvdz level of theory.}
	\label{tab:vpie_dz}
		\begin{tabular*}{\textwidth}{@{\extracolsep{\fill}}lc		
            d{1.3} d{1.3}} 
\toprule
            Compound & isotope        &  {\varepsilon}   &  \multicolumn{1}{c}{Exp.}     \\
			\midrule
			EtOH     & \ce{^13 C_{1}} &   -0.99          &  1.1\footnotemark[1]           \\
			EtOH     & \ce{^13 C_{2}} &    0.47          &  0.3\footnotemark[1]           \\
			MeOH     & \ce{^13 C}     &    0.93          &  0.5\footnotemark[1]           \\
\bottomrule
		\end{tabular*}
	\raggedright
	\footnotemark[1]{Ref. \citenum{Julien2015} by  distillation.} \\
\end{table*}
Contrary to experimental results, predicted VPIE are notoriously unaffected by the temperature change. In the case of \ce{^13 C_{2}} for EtOH and \ce{^13 C} in MeOH, predicted values are closer to experimental data, specially \ce{^13 C_{2}}. However, it must be recalled that it is precisely the position-specific isotope effect for carbon atom bonded \ce{OH}, the one that reflected large STD.

\begin{table*}
	\caption{VPIEs ($\varepsilon\enspace\permil$) for ethanol, and methanol, predicted at \(\omega\)B97X-V/aug-cc-pvtz level of theory.}
	\label{tab:vpie_tz}
		\begin{tabular*}{\textwidth}{@{\extracolsep{\fill}}lc		
            d{1.3} d{1.3}} 
\toprule
            Compound & isotope        &  {\varepsilon}   &  \multicolumn{1}{c}{Exp.}     \\
			\midrule
			EtOH     & \ce{^13 C_{1}} &   -0.94          &  1.1\footnotemark[1]           \\
			EtOH     & \ce{^13 C_{2}} &   -0.40          &  0.3\footnotemark[1]           \\
			MeOH     & \ce{^13 C}     &   -0.14          &  0.5\footnotemark[1]           \\
\bottomrule
		\end{tabular*}
	\raggedright
	\footnotemark[1]{Ref. \citenum{Julien2015} by  distillation.} \\
\end{table*}

\subsection{\label{sec:hessian} Effects from Hessian matrix approach}

\Cref{tab:par_vapie_dz} contains VIEs predictions for studies systems at \(\omega\)B97X-V/aug-cc-pvdz level of theory.
\begin{table*}
	\caption{VIEs ($\varepsilon\enspace\permil$) predicted from partial Hessian vibrational analysis at standard conditions for bromobenzene, trichlorofluoromethane, dibromomethane, ethanol, methanol, and trichloromethane, at \(\omega\)B97X-V/aug-cc-pvdz level of theory.}
	\label{tab:par_vapie_dz}
		\begin{tabular*}{\textwidth}{@{\extracolsep{\fill}}lc		
        d{1.3} r} %
\toprule
            Compound & isotope        &  {\varepsilon}   &  \multicolumn{1}{c}{Exp.}     \\
			\midrule
			BB       & \ce{^81 Br}    &    0.04          &{-0.9$\pm$0.2\footnotemark[1]} \\
			BB       & \ce{^13 C}     &    0.11          &{-0.4$\pm$0.1\footnotemark[1]} \\
			CFC      & \ce{^12 C}     &    0.63          & {1.7\footnotemark[2]} \\
			EtOH     & \ce{^13 C_{1}} &   -0.79          & -6.1\footnotemark[3]           \\
			EtOH     & \ce{^13 C_{2}} &    1.08          & -2.6\footnotemark[3]           \\
			MeOH     & \ce{^13 C}     &    0.87          & -6.5\footnotemark[3]           \\
			TCM      & \ce{^13 C}     &    0.19          &  1.2\footnotemark[2]           \\
\bottomrule
		\end{tabular*}
	\raggedright
	\footnotemark[1]{Ref. \citenum{Vasquez2018}.} \\
	\footnotemark[2]{Ref. \citenum{Horst2016}.}   \\
	\footnotemark[3]{Ref. \citenum{Julien2015} by passive evaporation.} \\
\end{table*}
Predominantly, numerical calculations reveal that unless molecules in the cluster are interacting directly \emph{via} hydrogen bond, frequencies are unaffected, hence allowing practically identical isotope effects as calculated with FHVA\@. Given result is highly valuable and promising, as it naturally allows us to accurately calculate with confidence \ce{^81 Br} and \ce{^13 C} isotope effects in BB, for which FHVA could not be computed owing to computational time limits. It is worth mentioning that computational time was decreased by 75 \%  in average by employing PHVA, as compared to FHVA.

It must be remembered that within the concept of the reduced partition function ratio (RPFR) by the Bigeleisen-Mayer approximation \cite{Bigeleisen1947}, it is equally possible to describe the isotope effect as the product of three factors, i) the product factor (PF), which contains molecular masses and moments of inertia, ii) the excitation factor (EXP), which introduces a correction for excited molecules, and iii) the zero-point energy contribution (ZPE) factor, which describes the zero-point energy differences for each of the vibrational modes \cite{Wolfsberg2009,VanHook2011}. As a rule, PF tends to remain a minor factor, as well as EXP which vanishes at elevated temperatures, therefore, ZPE represents the factor that typically determines the direction of the isotope effect (normal or inverse). 

Given that only one molecule in the cluster was active during PHVA, the PF in all performed calculations is close to one. The effect of PHVA in the remaining terms (EXP and ZPE) was studied by calculating those directly within isoC \cite{Vasquez2019}. For direct reference TCM and CFC from FHVA and PHVA were used, as precisely named systems did not present differences in the averaged VIE\@. There were not found relevant patterns for ZPE from PHVA\@. The only observed characteristic was that this term seemed to fluctuate in a convenient way to properly compensate the fact that EXC is always larger than its respective term from FHVA\@, thus ZPE values from PHVA were shifted. This observation is clearly evident when statistical parameters such as STD, skewness and kurtosis were compared between PHVA and FHVA, indicating close or identical values.

\subsection{\label{sec:sapt} Role of non-covalent interactions in the prediction of isotope effects}

Total interaction energies \emph{via} SAPT along with visualisation of non-covalent interactions from IGM analyses, was used in order characterise the magnitude of interactions in the first cluster of each compound optimised at the \(\omega\)B97X-V/aug-cc-pvdz level of theory (see \cref{fig:input_sapt_sup} in supplementary information). 

\begin{figure}
	\centering
	\begin{tikzpicture}
        \pgftext{%
	\includegraphics[width=\columnwidth]{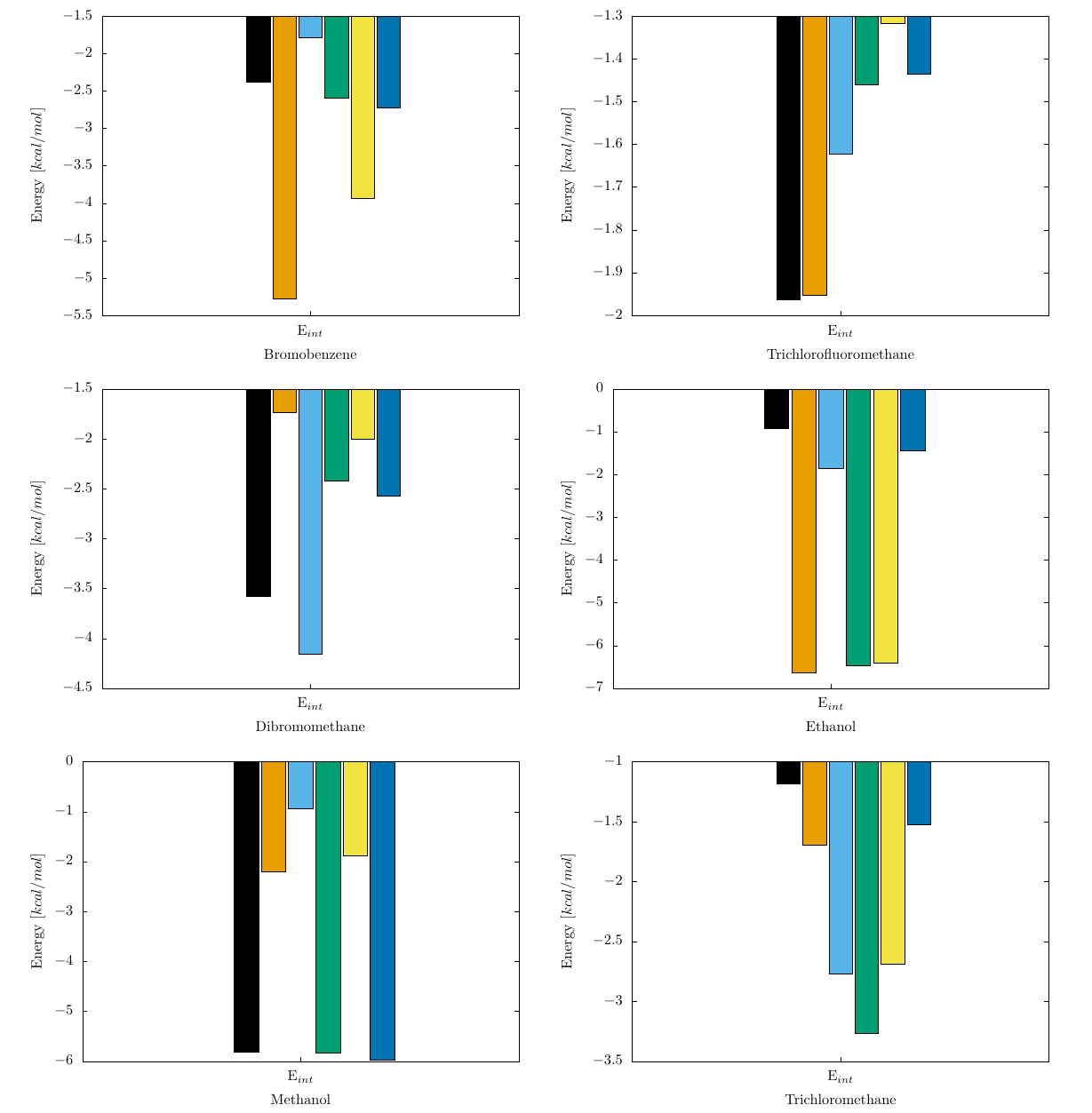}
	} ;
	\end{tikzpicture}
    \caption{Interaction energy for studied compounds of cluster one optimised at \(\omega\)B97x-V/aug-cc-pvdz. Each colour bar indicates one dimer combination (six per system), within a four molecule cluster.}
	\label{fig:sapt_int_dz}
\end{figure}

As each cluster consists precisely of four molecules, a total of six unique dimer arrangement can be formed, thus for instance to ethanol cluster the following dimers were formed: 1\(^{st}\) EtOH molecule with 2\(^{nd}\) EtOH molecule, 1\(^{st}\) EtOH molecule with 3\(^{rd}\) EtOH molecule, 1\(^{st}\) EtOH molecule with 4\(^{th}\) EtOH molecule, 2\(^{nd}\) EtOH molecule with 3\(^{rd}\) EtOH molecule, 2\(^{nd}\) EtOH molecule with 4\(^{th}\) EtOH molecule, and 3\(^{rd}\) EtOH molecule with 4\(^{th}\) EtOH molecule.
\Cref{fig:sapt_int_dz} shows bar plots with total interaction energies per dimer of each studied compound.

Results demonstrate that CFC possesses the weakest interactions among studied compounds. In increasing order, CFC is followed by TCM, DBM, BB, MeOH, and EtOH. Moreover, given order correlates with obtained reduced density gradients of intermolecular interactions from IGM analysis (see \cref{fig:igm_all} in supplementary information).

Although from SAPT calculations BB does not seem to typically reflect relatively strong interaction energies, the orange bar stands out over the rest. A comprehensive analysis of given result indicates it corresponds to a \(\pi\)-\(\pi\) stacking (parallel displaced), which is promptly confirmed after properly comparing dispersion and electrostatics components from the interaction energy decomposition (see \cref{fig:sapt_all_dz} in supplementary information).

The three bars (orange, green, and yellow) are prominent in EtOH SAPT analysis and precisely correspond to the formed hydrogen bonds. Average interaction energy of hydrogen bonds is 6.5 \si{\kilo\cal\per\mol}, which match perfectly Vargas-Caamal's average from high level of theory calculations \cite{VargasCaamal2015}. Remaining dimer interactions are relatively weak and principally raise from electrostatic interactions. As expected, MeOH interaction energies faithfully reflect a similar pattern with EtOH, \emph{i.e.} interaction mainly \emph{via} electrostatics, three dimers displaying potent attractive interactions (hydrogen bonding) and other three significantly weaker.

TCM interaction energies are the second weakest after CFC\@. Moreover, similar to CFC, BB and DBM intermolecular interactions are predominantly dispersion driven (see \cref{fig:sapt_all_dz} in supplementary information).

Among compounds DBM is genuinely interesting as SAPT analysis shows that it is the only compound among the studied ones, which interactions energies are proportionally distributed between electrostatics and dispersion (mixed type \cite{Rezac2011}). Furthermore, only in the case of DBM the induction component of the interaction energies is positive although weak in magnitude, therefore indicating a meagre DBM polarisation.

Other set of SAPT calculations were performed using optimised atomic positions of the fourth cluster at the \(\omega\)B97X-V/aug-cc-pvtz level of theory, in order to i) establish whether consistent patterns observed in the previously studied cluster prevail in other cluster, and ii) accurately identify general energetic changes with the increase of the basis set. Results (\cref{tab:avg_sapt}) suggest that in average minimal alterations in interactions energies are perceived.
\begin{table*}
    \caption{Average interaction energies from SAPT analysis for geometrically optimised clusters (BB, CFC, DBM, EtOH, MeOH, and TCM) at \(\omega\)B97X-V/aug-cc-pvdz and \(\omega\)B97X-V/aug-cc-pvdz levels of theory.}
	\label{tab:avg_sapt}
	\label{tab:par_vapie_dz}
		\begin{tabular*}{\textwidth}{@{\extracolsep{\fill}}l		
            d{1.3} d{1.3}}
\toprule
                     &                  \multicolumn{2}{c}{Ave. \(E_{\textrm{int}}\) [\si{\kilo\cal\per\mol}]}      \\
\cmidrule(lr){2-3}
            Compound &  \multicolumn{1}{c}{aug-cc-pvdz}  &  \multicolumn{1}{c}{aug-cc-pvtz}      \\
			\midrule
            BB         &      -3.12  &    -2.98  \\
			CFC        &      -1.63  &    -1.65  \\
            DBM        &      -2.74  &    -2.25  \\
			EtOH       &      -3.96  &    -3.85  \\
			MeOH       &      -3.77  &    -3.52  \\
			TCM        &      -2.19  &    -2.21  \\
\bottomrule
		\end{tabular*}
\end{table*}
Consequently, interaction energy components (electrostatics, exchange, induction, and dispersion) are seemingly unaffected (see \cref{fig:sapt_all_tz,fig:sapt_int_tz} in supplementary information). Plausible explanation to the dramatic transformation in the prediction of \ce{^81 Br} and \ce{^13 C} VIEs for DBM are sought in the SAPT results. Not surprisingly, DBM interaction energies are precisely the ones that comparatively change the most. We notice that distinctively one of the dimers does not possess any kind of intermolecular interaction, which results in an average weaker intermolecular interaction.

\begin{figure*}
	\centering
	\begin{tikzpicture}
	\pgftext{%
        \includegraphics[width=\columnwidth]{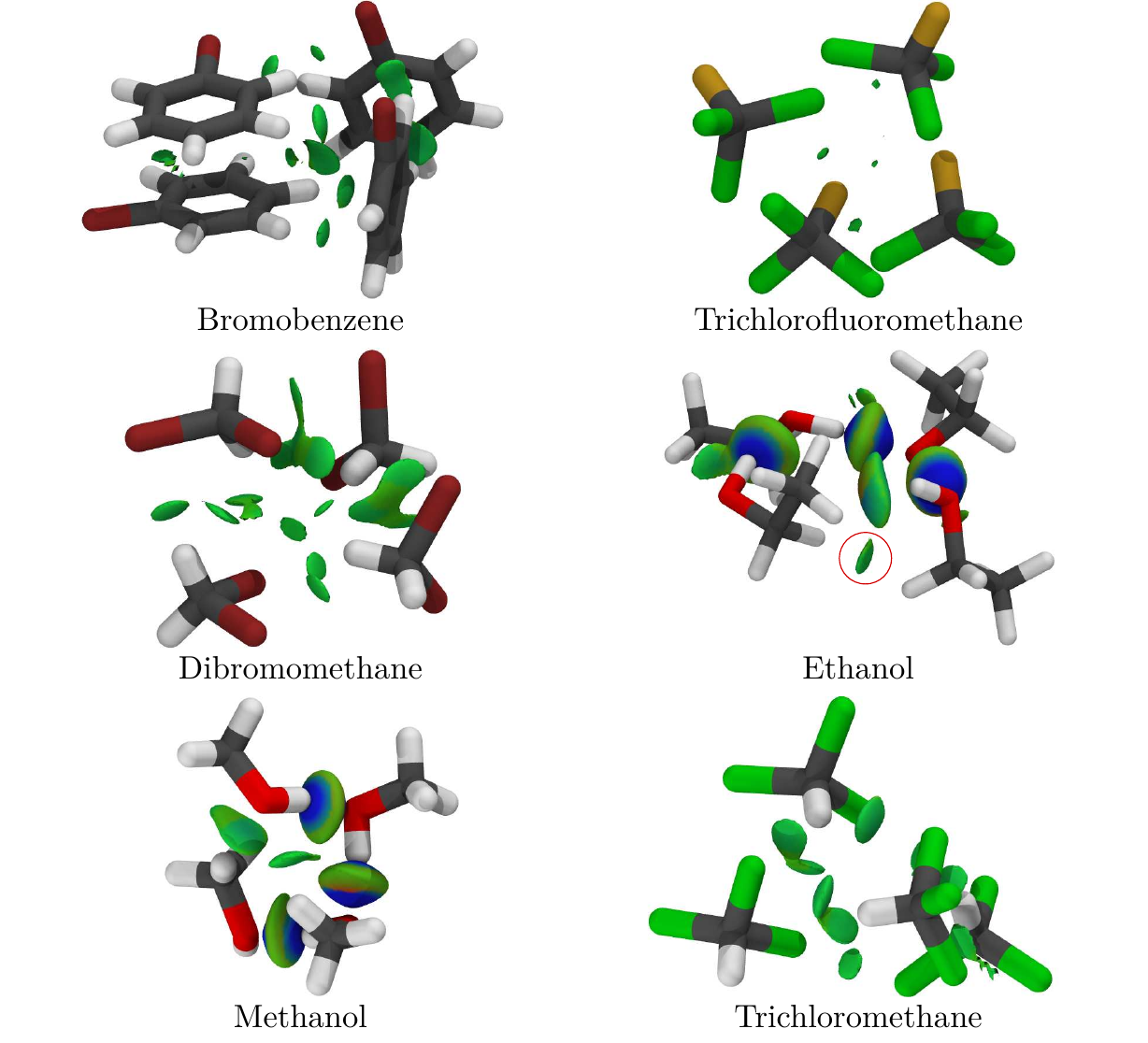}
	};
	\end{tikzpicture}
    \caption{Visualisation of intermolecular interactions from IGM analysis. BGR colour space describes intermolecular interactions, where strong attraction is highlighted in blue, van der Waals interactions are shown in green, and strong repulsion is displayed red.}
	\label{fig:igm_surf}
\end{figure*}

Visualisation of intermolecular interactions by IGM analysis are provided in \cref{fig:igm_surf}.
BGR colour space is used for representing characteristic interactions, thus blue implies strong attractive forces, green alludes van der Waals interactions, and red determines strong repulsive forces. Br atoms in BB do seem to occupy an active role in the interactions, which is most likely one of the possible reason why there were not VIEs in the PHVA\@. Overall weak interactions from SAPT analysis correlates with IGM picture of intermolecular interactions seen in CFC and TCM.

DBM demonstrates abundant interactions (van der Waals forces in green). However, this representation changes upon increasing the basis set. It is observed that the general extension of van der Waals forces is substantially diminished (see \cref{fig:igm_dbm_sub} in supplementary information), therefore enlarging the compelling evidence that points the weakening on interatomic interactions, as the key reason for the immoderate change of isotope effects prediction.

Using IGM for EtOH cluster, we detected that apparently a moderate van der Waals force (red circle in \cref{fig:igm_surf}) is responsible for the large STD during the calculation of \ce{^13 C_{2}} VIE\@. Given intermolecular interaction was only observed over/around \ce{C_{2}} atomic positions that, correspondingly, after being isotopically substituted to \ce{^{13} C}, exhibited relative large magnitudes (> 1.0\si{\permil}). Because EtOH molecules interact massively \emph{via} hydrogen bonds, it is unreasonable to dismiss the fact that the intensity or lack of hydrogen bonds might be equally contributing to the overall STD\@. Which is precisely the case of MeOH cluster. \ce{^13 C} isotope effects for MeOH display the largest STD among the studied compounds at the \(\omega\)B97X-V/aug-cc-pvdz level of theory, and there are not systematic patterns between position-specific isotope effects, SAPT, and IGM analysis that permit the proper identification of the large STD cause. Similarly IGM analysis on optimised atomic positions at the \(\omega\)B97X-V/aug-cc-pvtz level of theory (see \cref{fig:igm_met_sub} in supplementary information) does not provide straightforward insight into the issue.

\section{\label{sec:end} Conclusions}

We have calculated average volatilisation-pressure isotope effects for trichlorofluoromethane, dibromomethane, ethanol, methanol, and trichloromethane from full Hessian vibrational analysis at the \(\omega\)B97X-V/aug-cc-pvdz level of theory. Generally, isotope effects were predicted in qualitative agreement with experimental results. Isotope effects calculations with a larger basis set (aug-cc-pvtz) exhibit smaller standard deviation errors, and in most of the cases their magnitudes did not change appreciably compared to the smaller basis set (aug-cc-pvdz) results. Vapour-pressure isotope effects were in better agreement with experimental data. 

As for equilibrium isotope effects calculation, partial Hessian vibrational analysis \cite{Besley2008,Head1997} provides a robust, computational inexpensive and easy approach for their accurate calculation. We observed that unless intermolecular interactions of the studied compound are carried out via \emph{via} hydrogen bonds, accuracy of predicted isotope effects is uncompromised and equals isotope effect magnitudes from full Hessian vibrational analyses. It is substantial to note that using partial Hessian vibrational analysis, average computational time was reduced by a 75\% in comparison to full Hessian vibrational analysis.

Furthermore, we noticed that the product factor and the excitation factor of the reduced partition function ratio by the Bigeleisen-Mayer approximation \cite{Bigeleisen1947}, are almost one (> 0.999) in the vast majority of cases. Under those circumstances, the zero-point energy contribution from partial Hessian vibrational analysis shifts in order to equal predicted isotope effects from full Hessian vibrational analysis. With this in mind, we were able to predict \ce{^13 C} and \ce{^81 Br} isotope effects in bromobenzene, as it was not possible with full Hessian vibrational analysis due to computational time limitations.

The symmetry-adapted perturbation theory combined with independent gradient model has undoubtedly resulted in a successful scheme in order to appreciate qualitatively and quantitatively the consequences of intermolecular interactions in the calculation of equilibrium isotope effects. Thus for instance, the severe difference in the predicted values of \ce{^13 C} and \ce{^81 Br} isotope effects for dibromomethane when the size of basis set is increased, can be satisfactorily explained by the general reduction of intermolecular interactions. Although the present scheme was properly employed for studying all compounds, it is important to mention that owing to the relative strong interaction of hydrogen bonding systems (ethanol and methanol), comprehensible and straightforward assessment by independent gradient model becomes laborious. Because visually more interactions come in direct contact, which makes difficult to accurately distinguish interactions between specific atoms or regions. For the purpose of overcoming that inconvenience, we recommend to complementary use other more flexible visualisation techniques such as the electron localisation function (ELF) \cite{Becke1990} and atoms in molecules (AIM) \cite{KUMAR2016}. 

It is worthwhile to mention that the employed method for studying intermolecular interactions requires the use of high level of theory methods, or at least rigorous functionals within KS DFT, in order to reproduce compound atomic positions exact or relatively close to experimental data. Poor reproduction of atoms interaction in a cluster will undoubtedly lead to misleading interpretations of the PES, therefore we recommend well tested and benchmarked \cite{Goerigk2017} KS DFT functionals such as \(\omega\)B97X-V.

\section{Acknowledgments}
	This work was supported by the National
	Science Centre in Poland (Sonata BIS grant UMO-2014/14/E/ST4/00041)
	and in part by PLGrid Infrastructure (Poland). \\

\providecommand{\latin}[1]{#1}
\makeatletter
\providecommand{\doi}
  {\begingroup\let\do\@makeother\dospecials
  \catcode`\{=1 \catcode`\}=2 \doi@aux}
\providecommand{\doi@aux}[1]{\endgroup\texttt{#1}}
\makeatother
\providecommand*\mcitethebibliography{\thebibliography}
\csname @ifundefined\endcsname{endmcitethebibliography}
  {\let\endmcitethebibliography\endthebibliography}{}

\end{document}


\clearpage
\beginsupplement

\begin{table*}[htp!]
	\caption{VPIEs ($\varepsilon\enspace\permil$) predicted on evaporation using aug-cc-pvdz basis set for 
    trichlorofluoromethane, dibromomethane, ethanol, methanol, and trichloromethane.}
	\label{tab:vapie_dz_sup}
		\begin{tabular*}{\textwidth}{@{\extracolsep{\fill}}l d{1.3}		
            d{1.3} d{1.3} d{1.3}} 
\toprule
            Compound             & isotope        & \multicolumn{1}{c}{$\varepsilon$} & \multicolumn{1}{c}{STD}    &  \multicolumn{1}{c}{Exp.}     \\
			\midrule
			CFC                  &  \ce{^12 C}    &       0.6362  &  0.1005   & 1.7\footnotemark[1] \\
			DBM                  & \ce{^81 Br}   &      -1.7366  &  0.0329   & -0.8	0.1\footnotemark[2] \\
            DBM                  & \ce{^13 C}     &      -1.9926  &  0.2431   &  0.6\footnotemark[2]          \\
			EtOH                 & \ce{^13 C_{1}} &      -1.2557  &  0.6583   & -6.1\footnotemark[3]           \\
			EtOH                 & \ce{^13 C_{2}} &       0.5646  &  0.5798   & -2.6\footnotemark[3]           \\
			EtOH\footnotemark[4] & \ce{^13 C_{2}} &       0.2638  &  0.2402   & -2.6\footnotemark[3]           \\
			EtOH\footnotemark[5] & \ce{^13 C_{2}} &       1.1469  &  0.5603   & -2.6\footnotemark[3]           \\
			MeOH                 & \ce{^13 C}     &       1.0408  &  3.9015   & -6.5\footnotemark[3]           \\
			TCM                  & \ce{^13 C}     &       0.1618  &  0.2890   &  1.2\footnotemark[1]           \\
\bottomrule		
\end{tabular*}
	\raggedright
	\footnotemark[1]{Ref. \citenum{Horst2016}.}   \\
	\footnotemark[2]{Ref. \citenum{Vasquez2018}.} \\
	\footnotemark[3]{Ref. \citenum{Julien2015} by passive evaporation.}      \\
	\footnotemark[4]{Considering position-specific isotope effects smaller than one.}      \\
	\footnotemark[5]{Considering position-specific isotope effects larger than one.}      \\
\end{table*}                                                                

\begin{table*}[htp!]
	\caption{VPIEs ($\varepsilon\enspace\permil$) predicted on evaporation using aug-cc-pvtz basis set for 
    trichlorofluoromethane, dibromomethane, ethanol, methanol, and trichloromethane.}
	\label{tab:vapie_tz_sup}
		\begin{tabular*}{\textwidth}{@{\extracolsep{\fill}}lc		
            d{1.3} d{1.3} d{1.3}} 
\toprule
            Compound             & isotope        & \multicolumn{1}{c}{$\varepsilon$} & \multicolumn{1}{c}{STD}    &  \multicolumn{1}{c}{Exp.}     \\
			\midrule
			CFC                  & \ce{^12 C}     &    0.6429     &  0.1901   & 1.7\footnotemark[1] \\
			DBM                  & \ce{^81 Br}    &   -0.0206     &  0.0268   &-0.8\footnotemark[2] \\
            DBM                  & \ce{^13 C}     &   -0.1737     &  0.1921   &  0.6\footnotemark[2]         \\
			EtOH                 & \ce{^13 C_{1}} &   -1.1846     &  0.6461   & -6.1\footnotemark[3]           \\
			EtOH                 & \ce{^13 C_{2}} &    0.4718     &  0.5618   & -2.6\footnotemark[3]           \\
			EtOH\footnotemark[4] & \ce{^13 C_{2}} &    0.1734     &  0.1986   & -2.6\footnotemark[3]           \\
			EtOH\footnotemark[5] & \ce{^13 C_{2}} &    1.3672     &  0.2746   & -2.6\footnotemark[3]           \\
			MeOH                 & \ce{^13 C}     &   -0.1250     &  0.4248   & -6.5\footnotemark[3]           \\
			TCM                  & \ce{^13 C}     &    0.0920     &  0.2152   &  1.2\footnotemark[1]           \\
\bottomrule		
\end{tabular*}
	\raggedright
	\footnotemark[1]{Ref. \citenum{Horst2016}.}   \\
	\footnotemark[2]{Ref. \citenum{Vasquez2018}.} \\
	\footnotemark[3]{Ref. \citenum{Julien2015} by passive evaporation.}      \\
	\footnotemark[4]{Considering position-specific isotope effects smaller than one.}      \\
	\footnotemark[5]{Considering position-specific isotope effects larger than one.}      \\
\end{table*}                                                                

\begin{table*}[htp!]
	\caption{VPIEs ($\varepsilon\enspace\permil$) predicted on evaporation using aug-cc-pvtz basis set for 
    trichlorofluoromethane, dibromomethane, ethanol, methanol, and trichloromethane.}
	\label{tab:vpie_tz_sub}
		\begin{tabular*}{\textwidth}{@{\extracolsep{\fill}}lc		
            d{1.3} d{1.3} d{1.3}} 
\toprule
            Compound             & isotope        & \multicolumn{1}{c}{$\varepsilon$} & \multicolumn{1}{c}{STD}    &  \multicolumn{1}{c}{Exp.}     \\
			\midrule
			EtOH                 & \ce{^13 C_{1}} &    -0.9927    &  0.5353   & 1.1\footnotemark[1]           \\
			EtOH                 & \ce{^13 C_{2}} &     0.4694    &  0.4530   & 0.3\footnotemark[1]           \\
			EtOH\footnotemark[2] & \ce{^13 C_{2}} &     0.2355    &  0.1887   & 0.4\footnotemark[1]           \\
			EtOH\footnotemark[3] & \ce{^13 C_{2}} &     1.1710    &  0.2403   & 0.3\footnotemark[1]           \\
			MeOH                 & \ce{^13 C}     &     0.9342    &  3.2759   & 0.5\footnotemark[1]           \\
\bottomrule		
\end{tabular*}
	\raggedright
	\footnotemark[1]{Ref. \citenum{Julien2015} by passive evaporation.}      \\
	\footnotemark[2]{Considering position-specific isotope effects smaller than one.}      \\
	\footnotemark[3]{Considering position-specific isotope effects larger than one.}      \\
\end{table*}                                                                

\begin{figure}[htp!]
	\setlength\tabcolsep{1pt}
	\settowidth\rotheadsize{Radcliffe Cam}
	\begin{tabularx}{\linewidth}{XX}
		   \includegraphics[width=\hsize,valign=m]{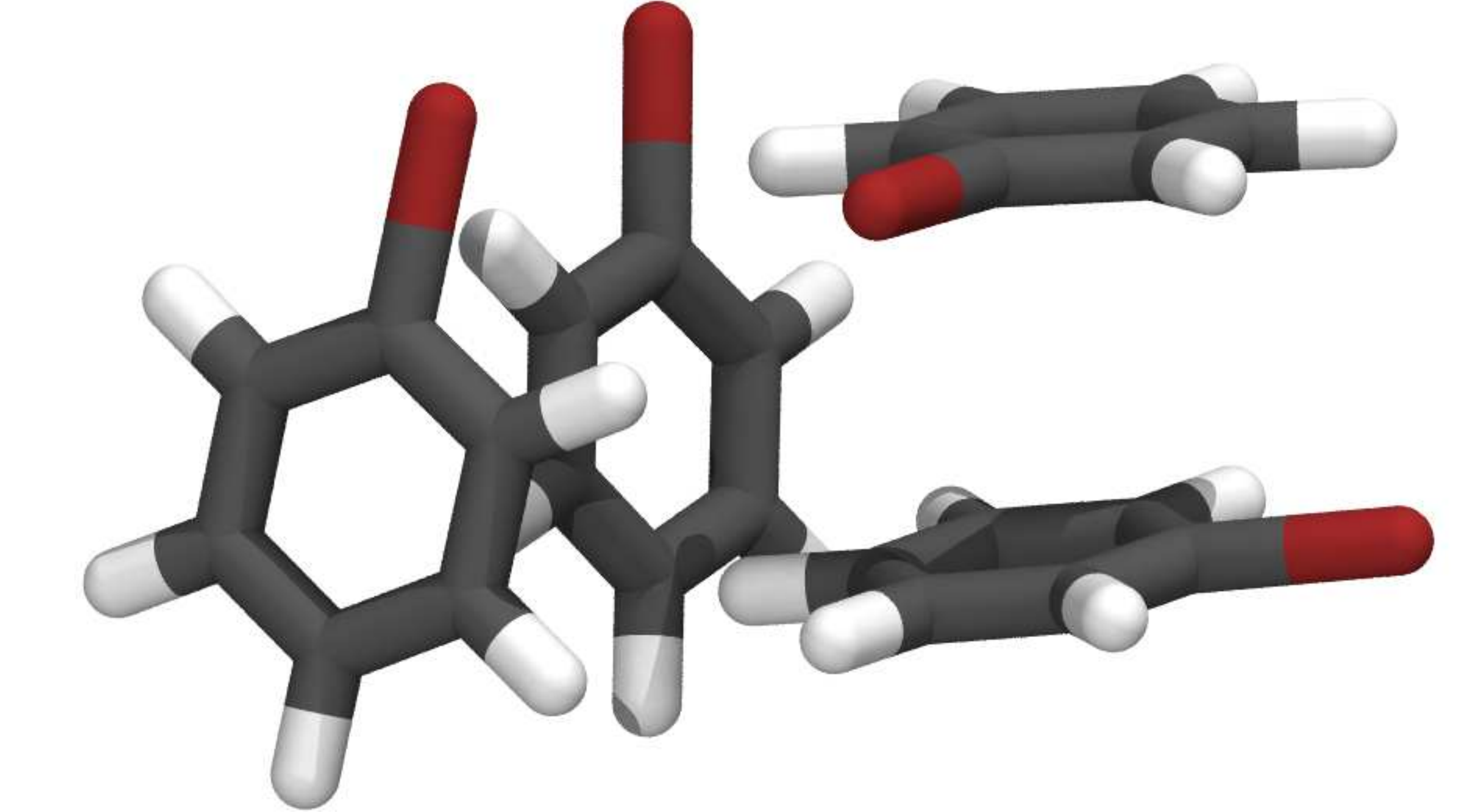}
		 & \includegraphics[width=\hsize,valign=m]{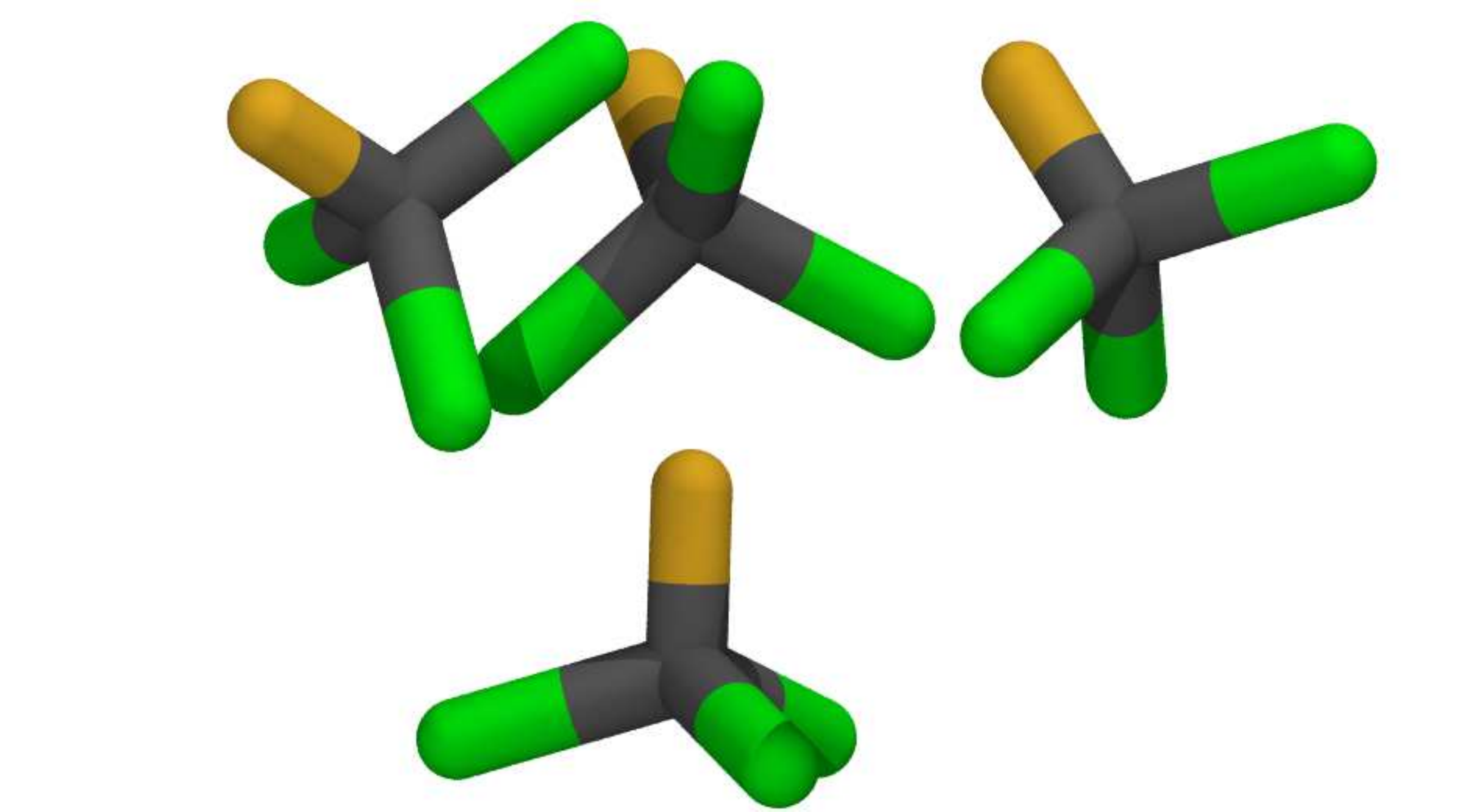}  \\  \addlinespace[1pt]
           \multicolumn{1}{c}{Bromobenzene} & \multicolumn{1}{c}{Trichlorofluoromethane}  \\  \addlinespace[2pt]
		   \includegraphics[width=\hsize,valign=m]{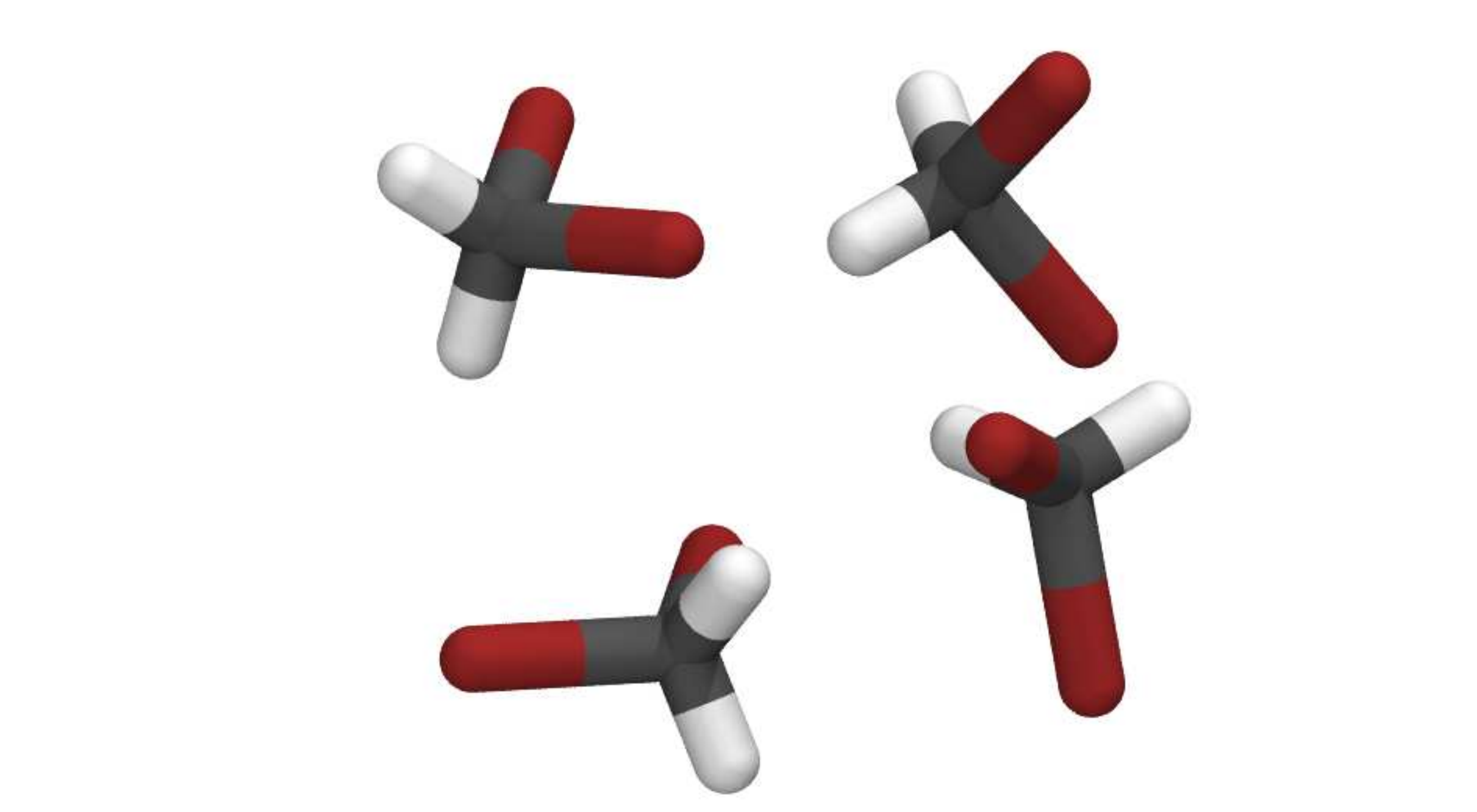}
		 & \includegraphics[width=\hsize,valign=m]{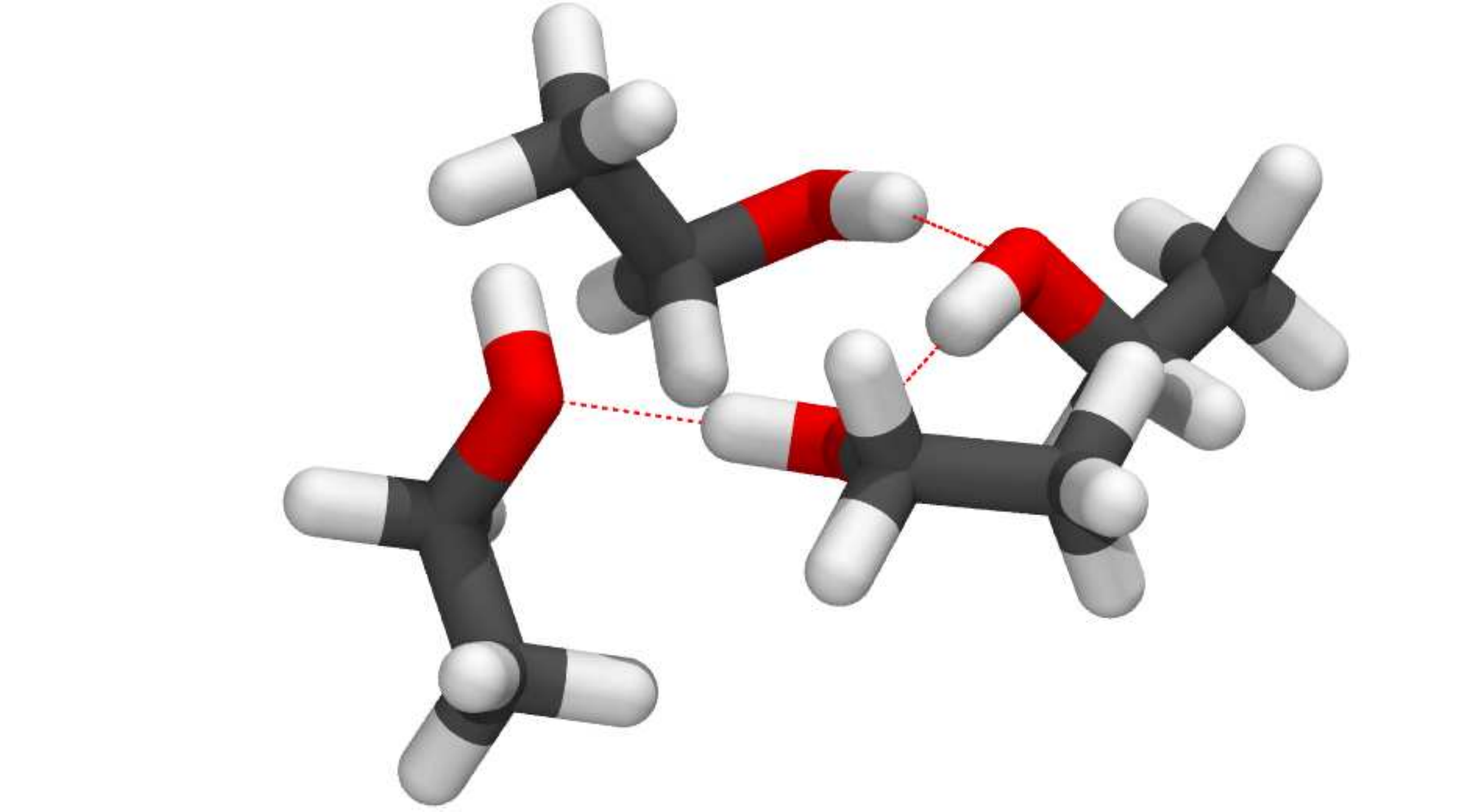} \\  \addlinespace[1pt]
           \multicolumn{1}{c}{Dibromomethane} & \multicolumn{1}{c}{Ethanol}  \\  \addlinespace[2pt]
		   \includegraphics[width=\hsize,valign=m]{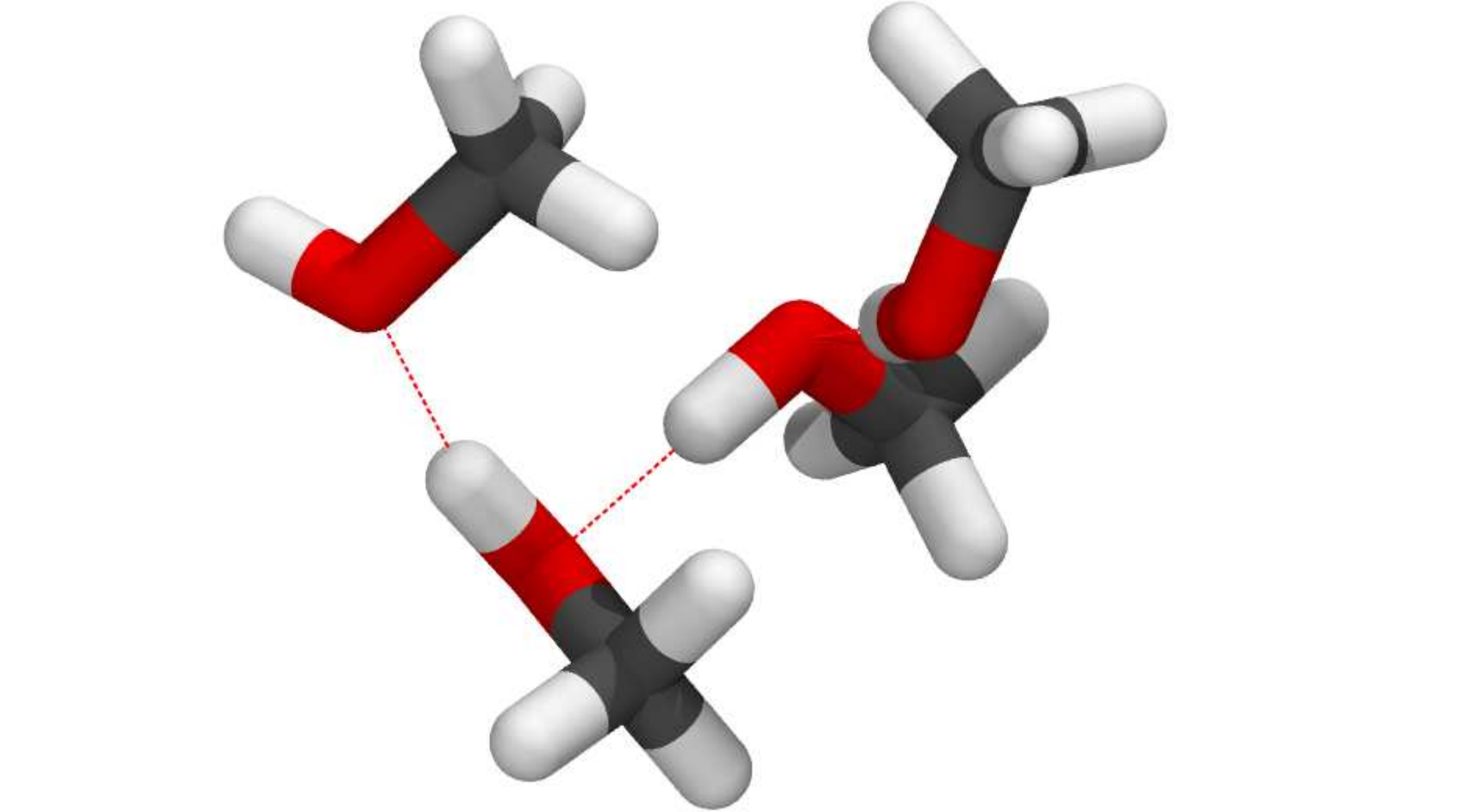}
		 & \includegraphics[width=\hsize,valign=m]{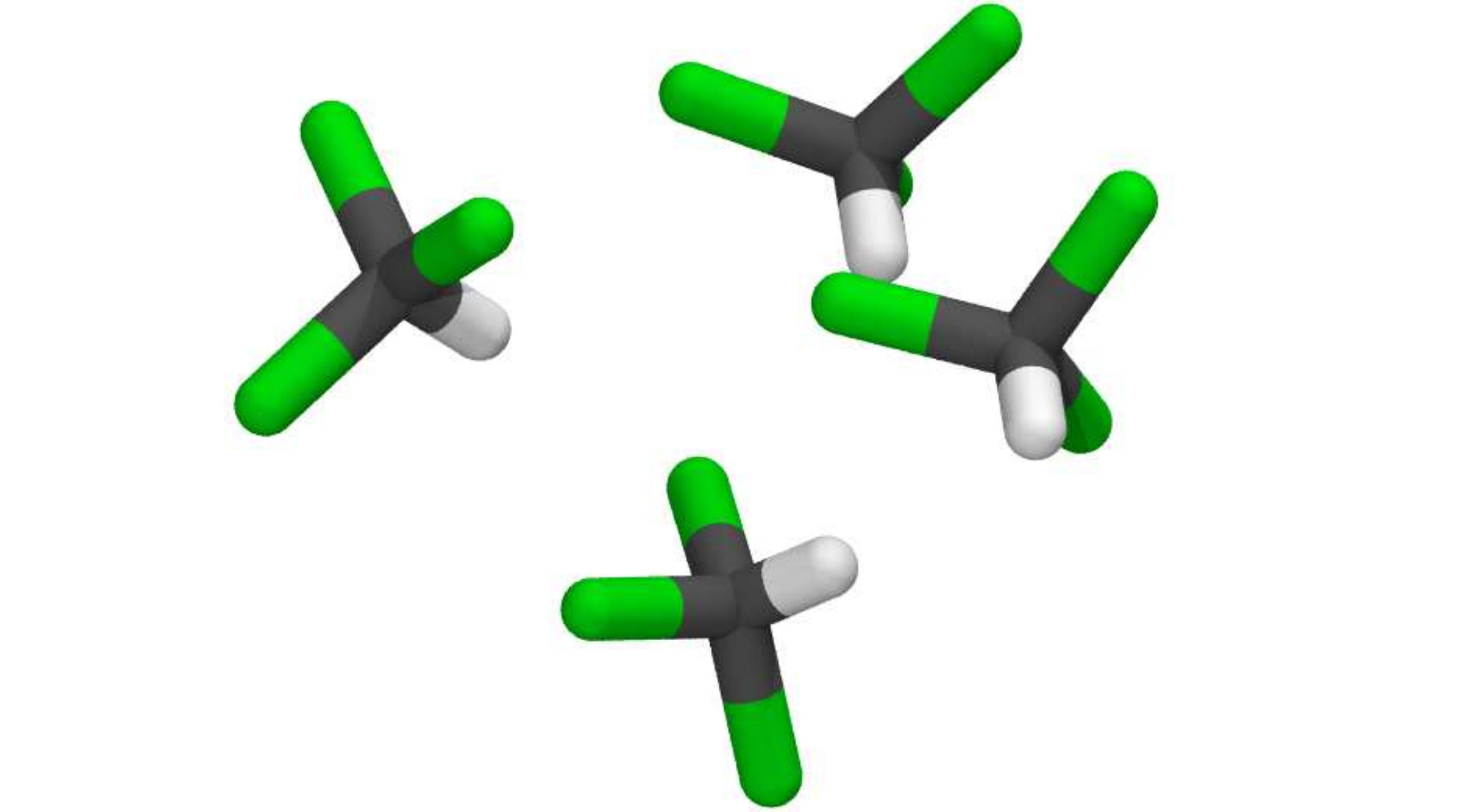} \\  \addlinespace[1pt]
           \multicolumn{1}{c}{Methanol} & \multicolumn{1}{c}{Trichloromethane}  \\  \addlinespace[1pt]
	\end{tabularx}
	\caption{Atomic positions visualisation of compounds cluster used for SAPT and IGM analysis.}
	\label{fig:input_sapt_sup}
\end{figure}

\begin{figure*}[htp!]
	\centering
	\begin{tikzpicture}
        \pgftext{%
	\includegraphics[width=1\columnwidth]{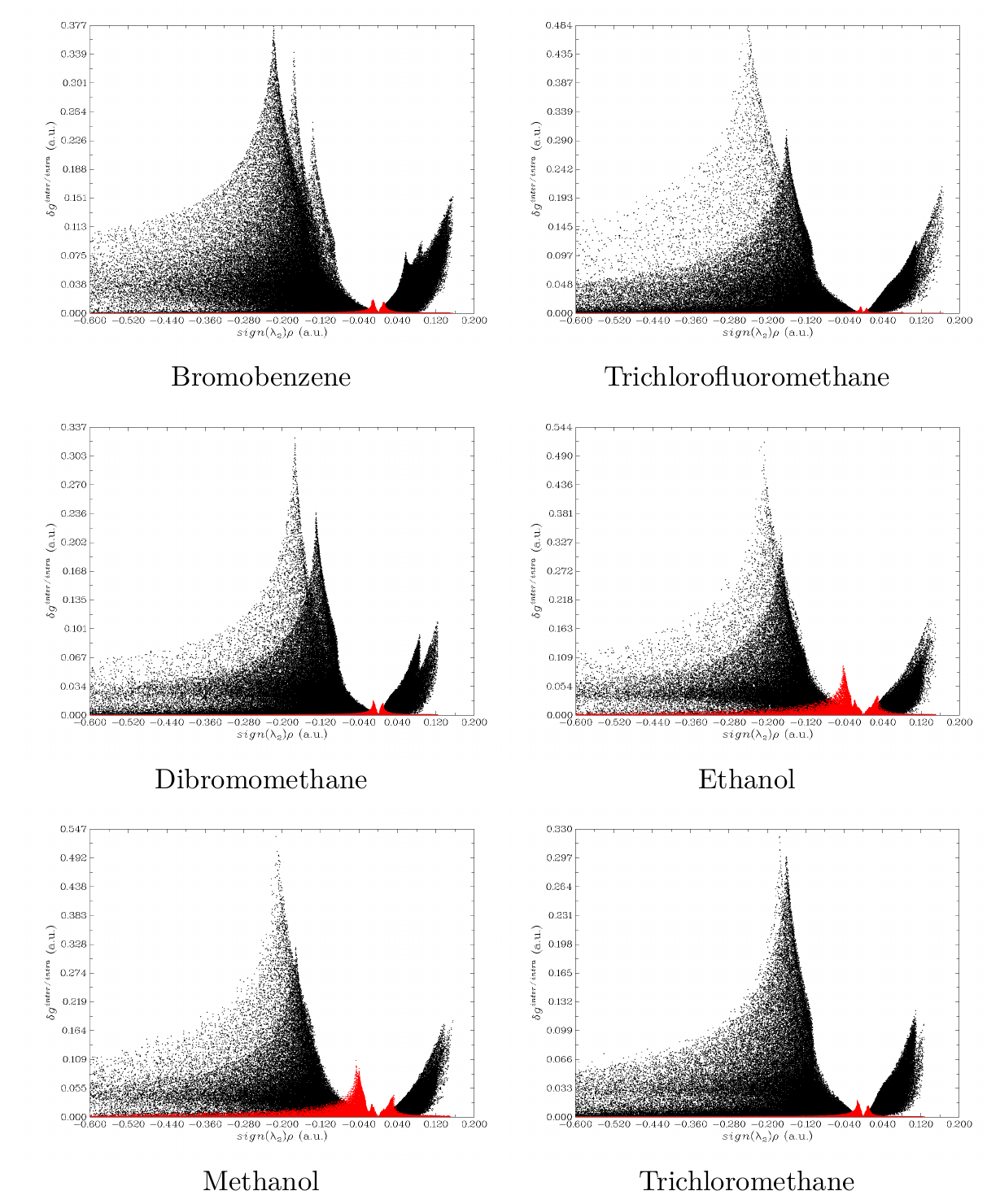}
	} ;
	\end{tikzpicture}
    \caption{Scattered plot of intermolecular (red) and intramolecular (black) interactions from IGM analysis.}
	\label{fig:igm_all}
\end{figure*}

\begin{figure*}[htp!]
	\centering
	\begin{tikzpicture}
        \pgftext{%
	\includegraphics[width=\columnwidth]{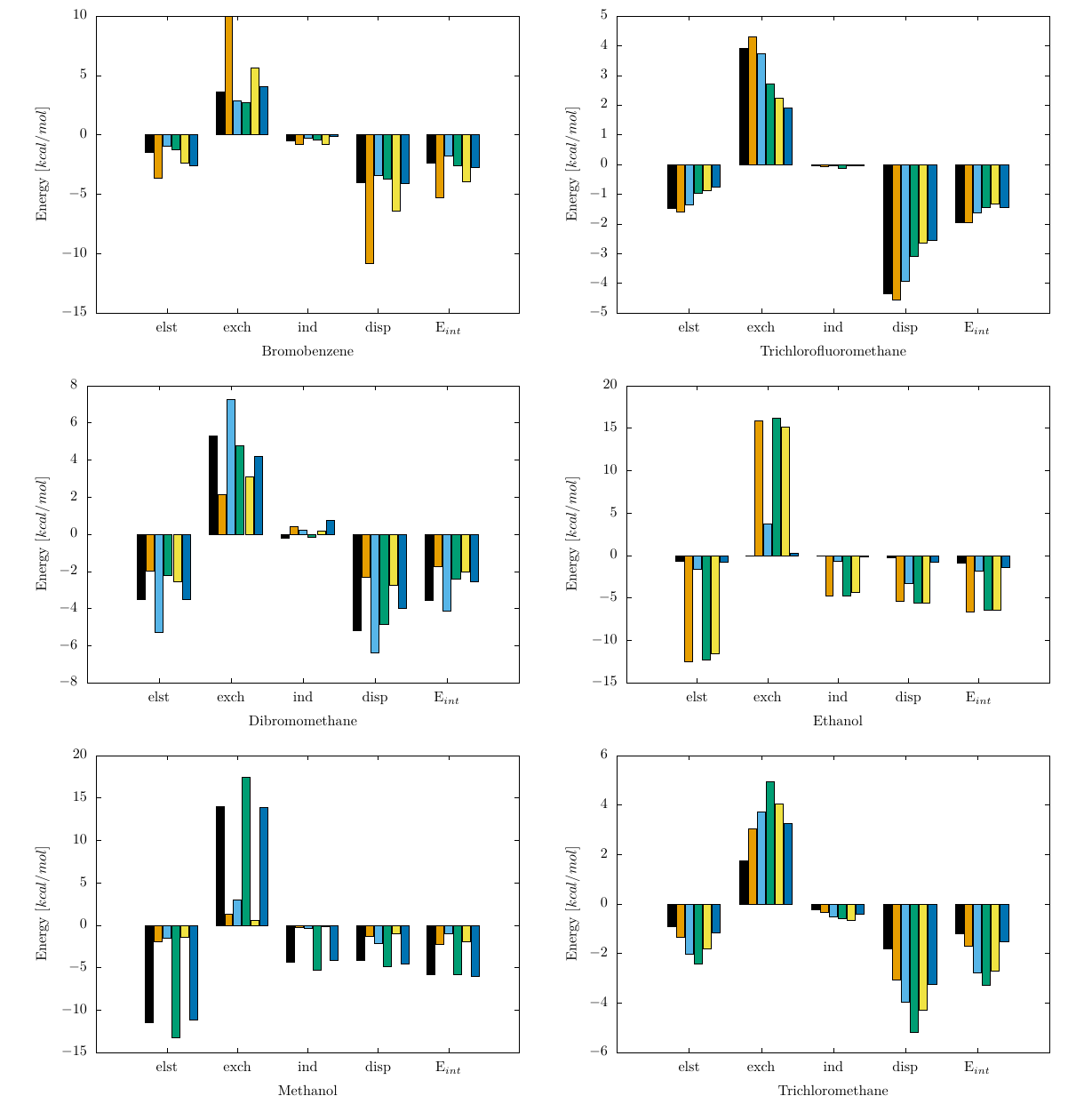}
	} ;
	\end{tikzpicture}
    \caption{Interaction energy and its decomposition into electrostatics (elst), exchange (exch), induction (ind), and dispersion (disp) for studied compounds of cluster one optimised at \(\omega\)B97x-V/aug-cc-pvdz. Each colour bar indicates one dimer combination, within a four molecule cluster.}
	\label{fig:sapt_all_dz}
\end{figure*}

\begin{figure*}[htp!]
	\centering
	\begin{tikzpicture}
        \pgftext{%
	\includegraphics[width=\columnwidth]{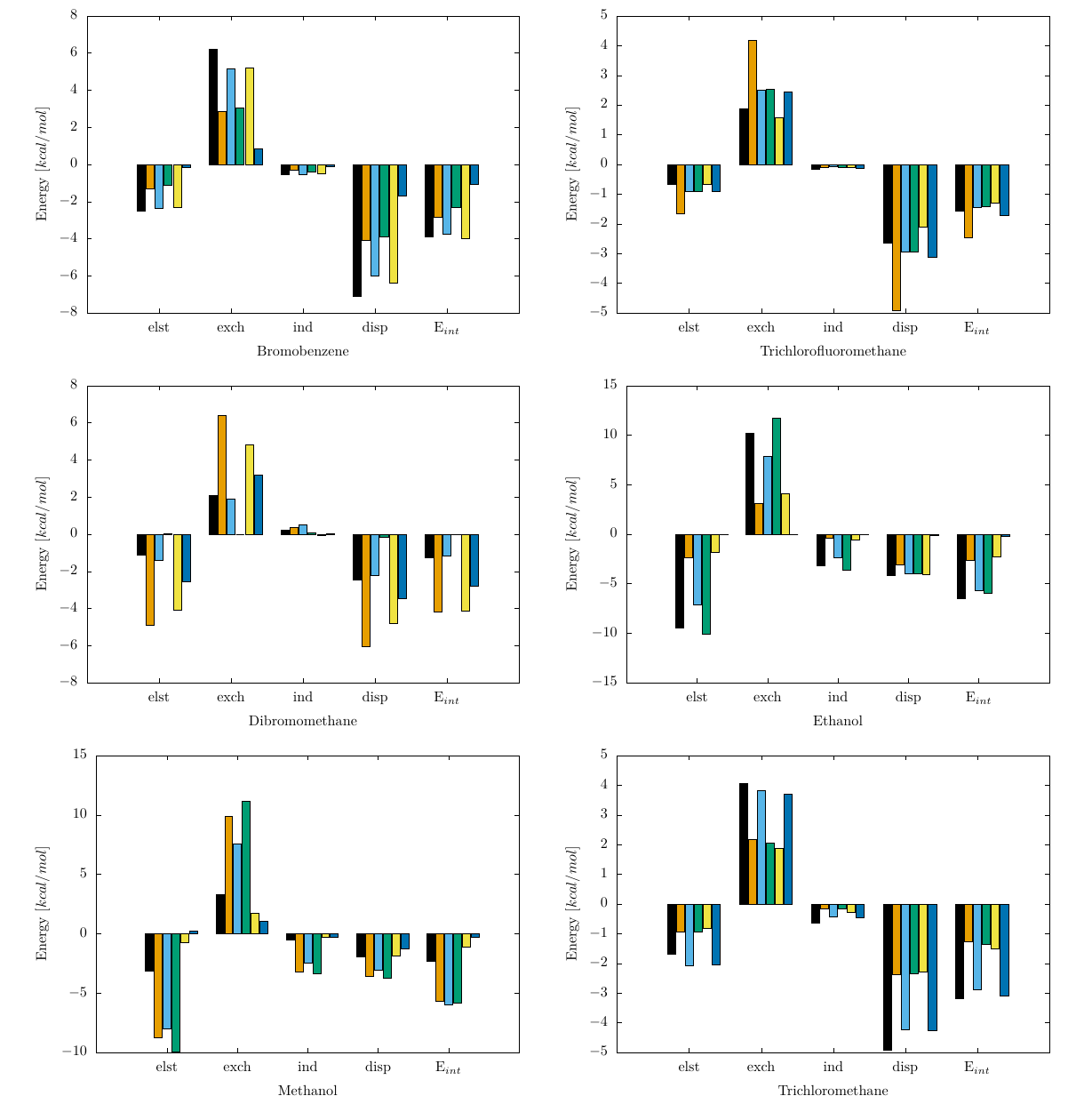}
	} ;
	\end{tikzpicture}
    \caption{Interaction energy and its decomposition into electrostatics (elst), exchange (exch), induction (ind), and dispersion (disp) for studied compounds of cluster one optimised at \(\omega\)B97x-V/aug-cc-pvdz. Each colour bar indicates one dimer combination, within a four molecule cluster.}
	\label{fig:sapt_all_tz}
\end{figure*}

\begin{figure*}[htp!]
	\centering
	\begin{tikzpicture}
        \pgftext{%
	\includegraphics[width=\columnwidth]{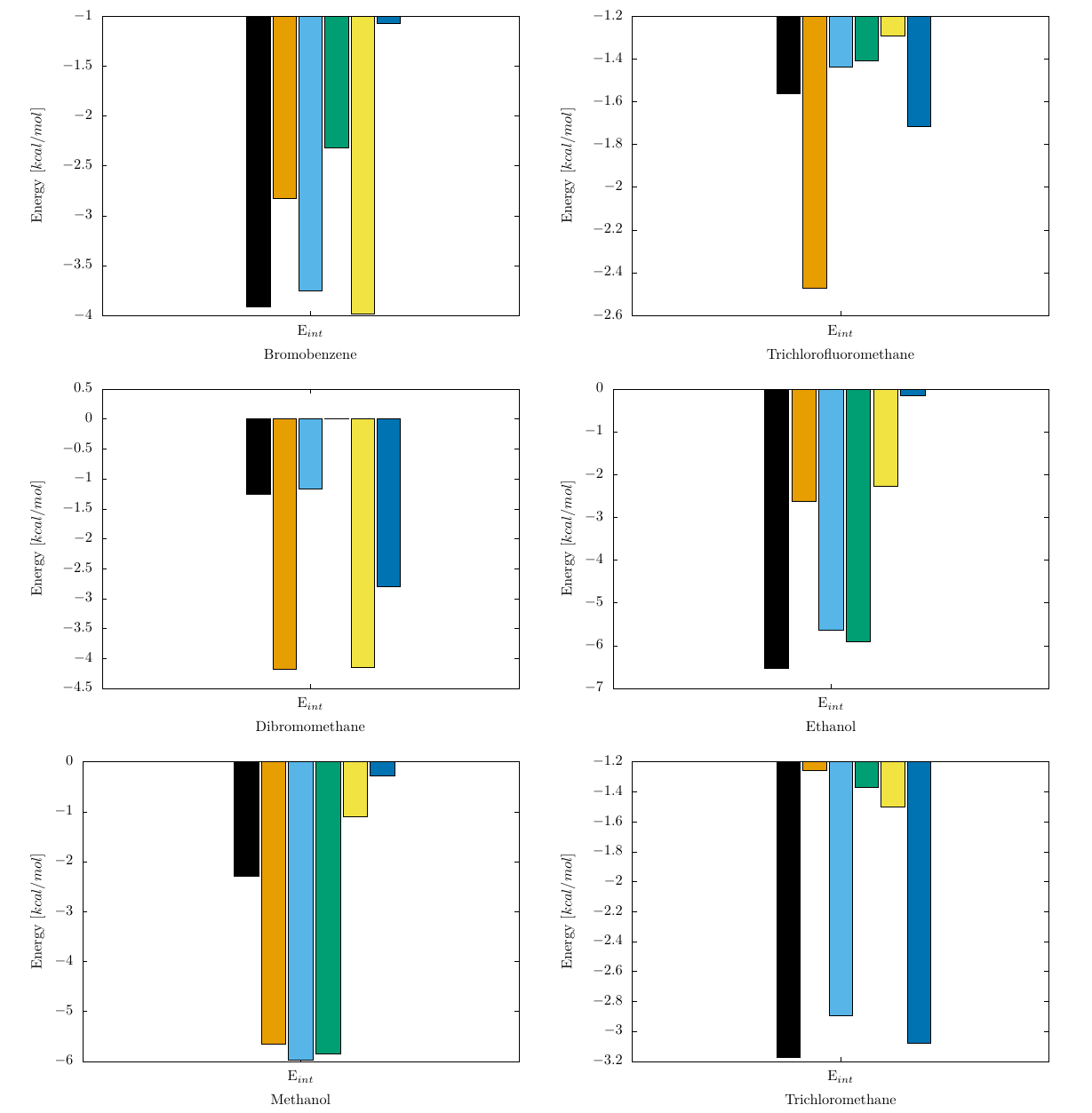}
	} ;
	\end{tikzpicture}
    \caption{Interaction energy for studied compounds of cluster one optimised at \(\omega\)B97x-V/aug-cc-pvtz. Each colour bar indicates one dimer combination, within a four molecule cluster.}
	\label{fig:sapt_int_tz}
\end{figure*}

\begin{figure}[htp!]
	\setlength\tabcolsep{1pt}
	\settowidth\rotheadsize{Radcliffe Cam}
	\begin{tabularx}{\linewidth}{XX}
		   \includegraphics[width=\hsize,valign=m]{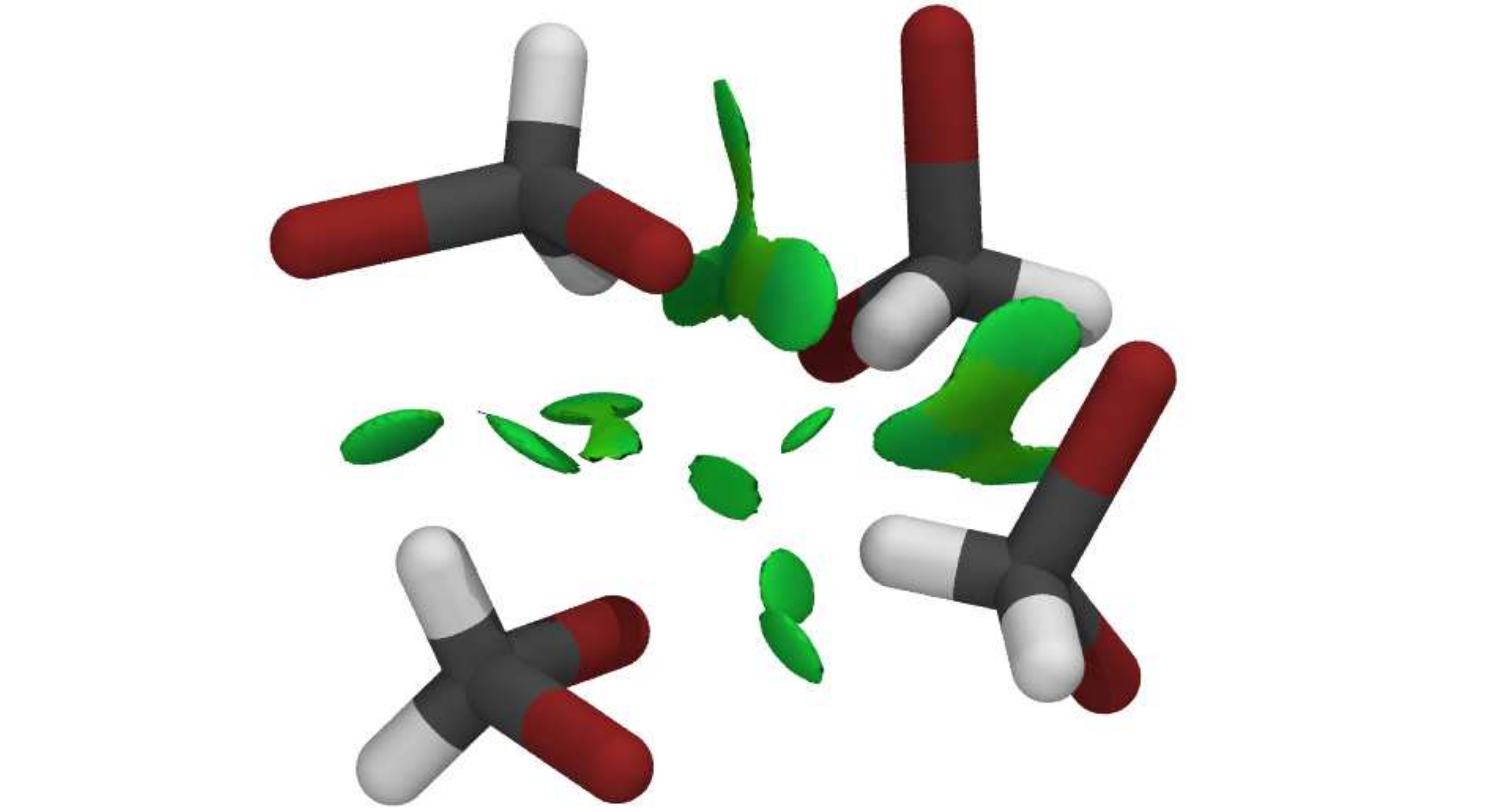}
		 & \includegraphics[width=\hsize,valign=m]{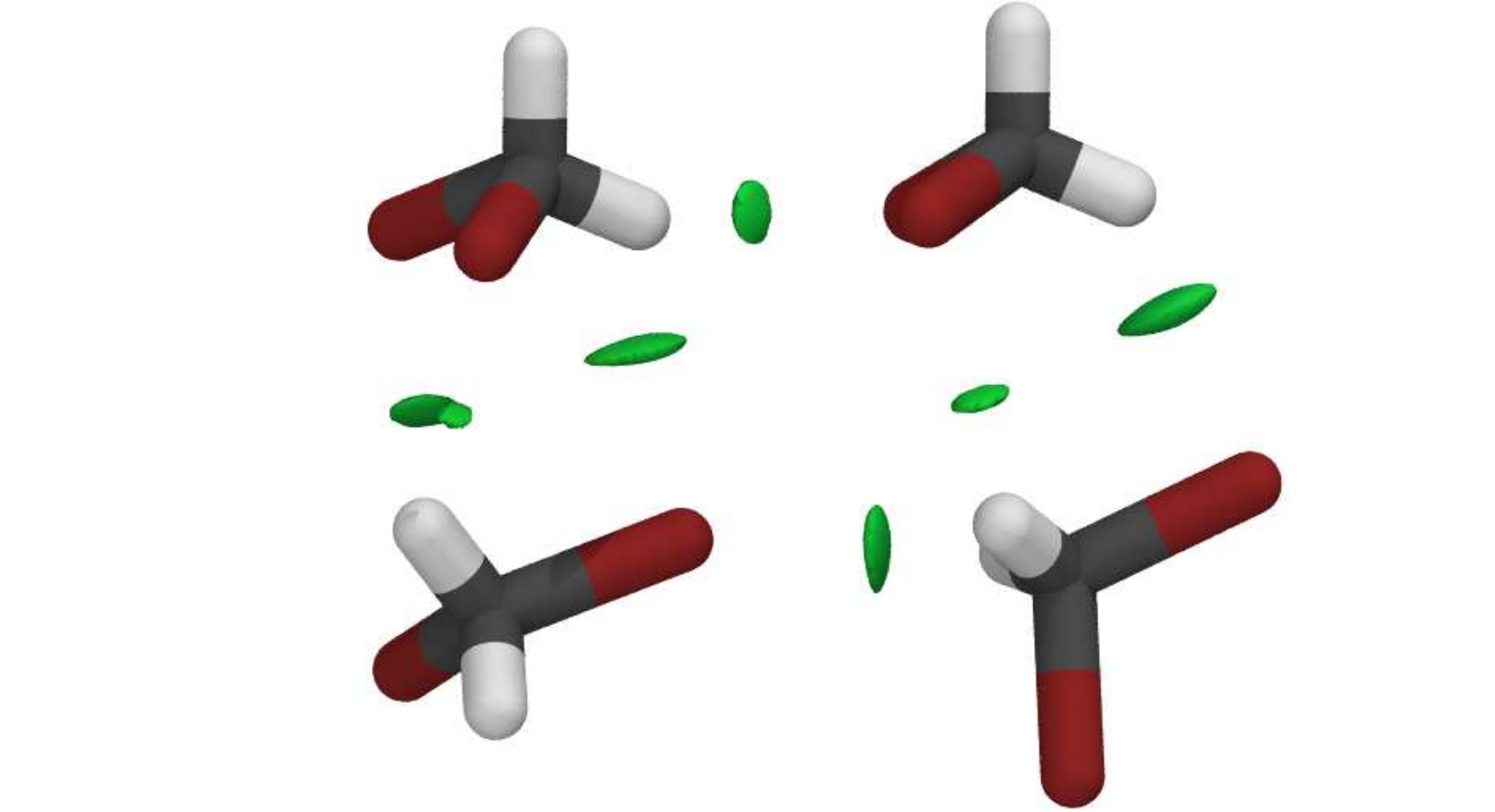}  \\  \addlinespace[1pt]
        \multicolumn{1}{c}{\(\omega\)B97x-V/aug-cc-pvdz} & \multicolumn{1}{c}{\(\omega\)B97x-V/aug-cc-pvtz}  \\  \addlinespace[1pt]
	\end{tabularx}
    \caption{Visualisation of DBM intermolecular interactions by IGM analysis to optimised atomic positions at the \(\omega\)B97x-V/aug-cc-pvdz (left) and \(\omega\)B97x-V/aug-cc-pvdz (rigth) level of theory. BGR colour space describes intermolecular interactions, where blue highlights strong attraction, green are van der Waals interactions, and red are strong repulsive.}
	\label{fig:igm_dbm_sub}
\end{figure}

\begin{figure}[htp!]
	\setlength\tabcolsep{1pt}
	\settowidth\rotheadsize{Radcliffe Cam}
	\begin{tabularx}{\linewidth}{XX}
		   \includegraphics[width=\hsize,valign=m]{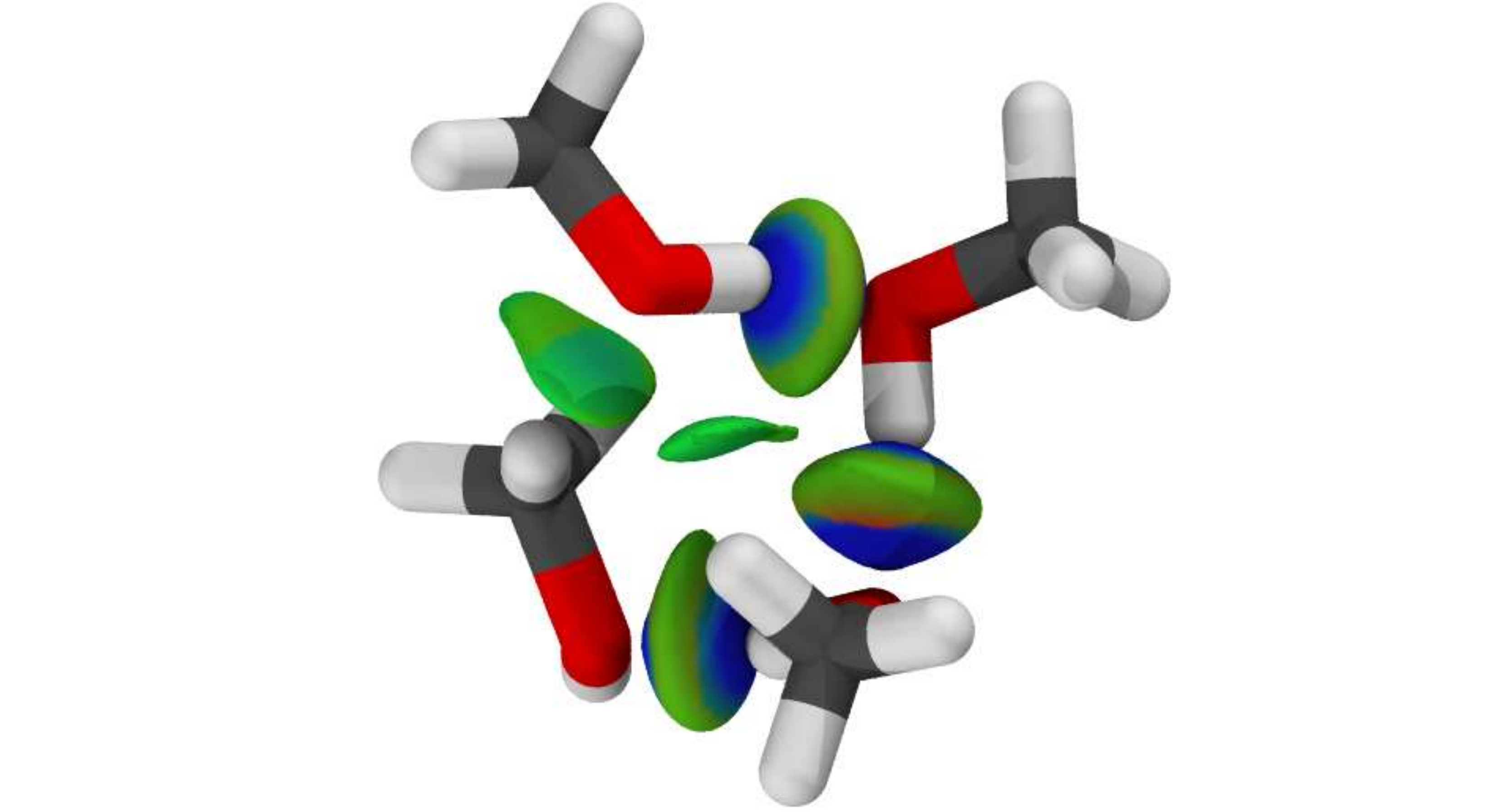}
		 & \includegraphics[width=\hsize,valign=m]{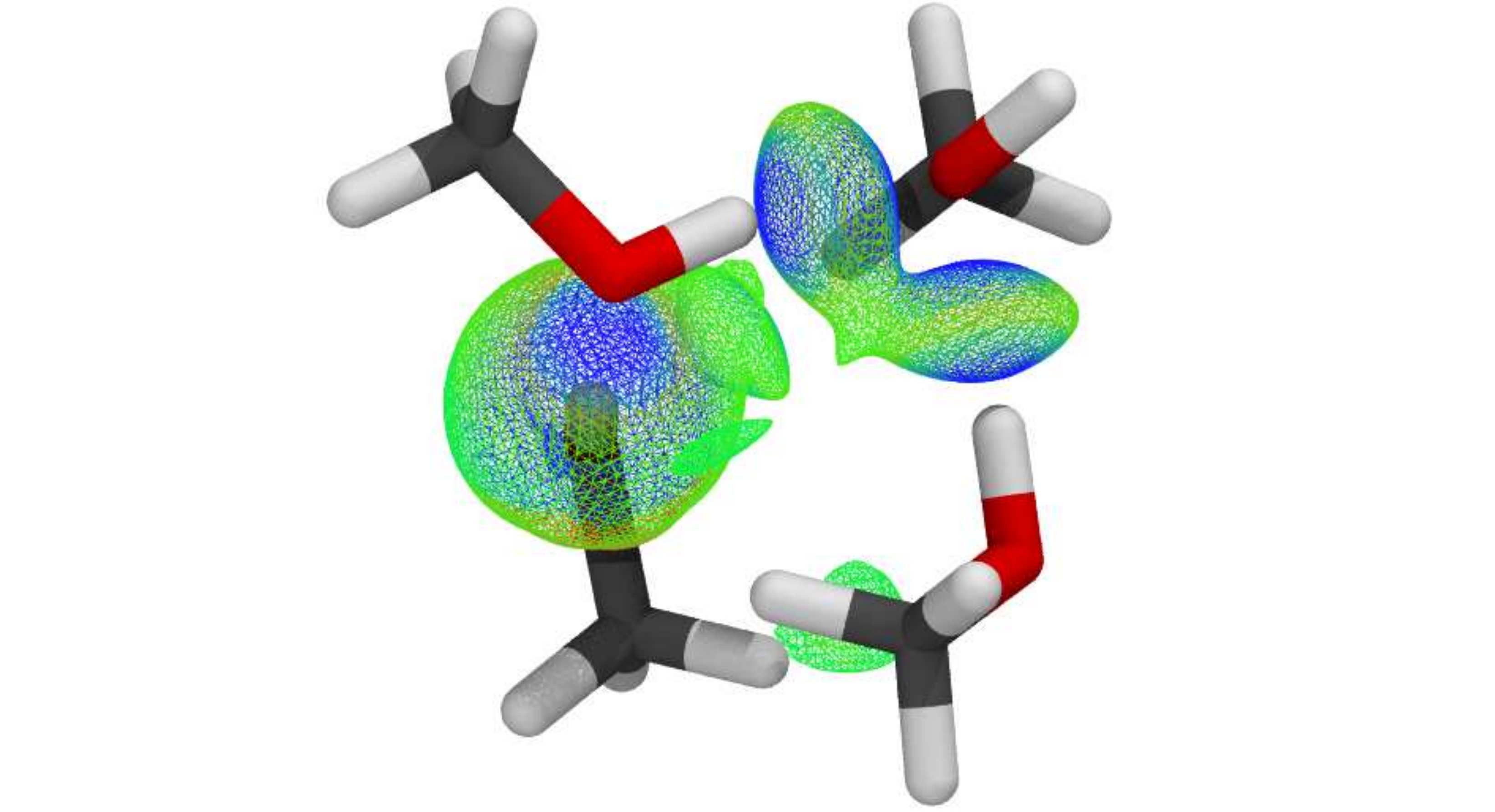}  \\  \addlinespace[1pt]
        \multicolumn{1}{c}{\(\omega\)B97x-V/aug-cc-pvdz} & \multicolumn{1}{c}{\(\omega\)B97x-V/aug-cc-pvtz}  \\  \addlinespace[1pt]
	\end{tabularx}
    \caption{Visualisation of MeOH intermolecular interactions by IGM analysis to optimised atomic positions at the \(\omega\)B97x-V/aug-cc-pvdz (left) and \(\omega\)B97x-V/aug-cc-pvdz (rigth) level of theory. BGR colour space describes intermolecular interactions, where blue highlights strong attraction, green are van der Waals interactions, and red are strong repulsive.}
	\label{fig:igm_met_sub}
\end{figure}

\providecommand{\latin}[1]{#1}
\makeatletter
\providecommand{\doi}
  {\begingroup\let\do\@makeother\dospecials
  \catcode`\{=1 \catcode`\}=2 \doi@aux}
\providecommand{\doi@aux}[1]{\endgroup\texttt{#1}}
\makeatother
\providecommand*\mcitethebibliography{\thebibliography}
\csname @ifundefined\endcsname{endmcitethebibliography}
  {\let\endmcitethebibliography\endthebibliography}{}